\documentclass[aps,pra,twocolumn,groupedaddress,showpacs,showkeys,amsfonts]{revtex4-1}
\usepackage{amsfonts}
\usepackage{amsmath}
\usepackage{amssymb}
\usepackage{graphicx}
\usepackage{epsf}
\usepackage{epsfig}
\usepackage{hyperref}
\usepackage{tabularx}
\newcommand{\eq}[1]{Eq.~\eqref{#1}}
\newcommand{\eqs}[1]{Eqs.~\eqref{#1}}
\newcommand{\seq}[1]{Sec.~\ref{#1}}
\newcommand{\app}[1]{App.~\ref{#1}}
\newcommand{\fig}[1]{Fig.~\ref{#1}}
\newcommand{\be}{\begin{equation}}
\newcommand{\ee}{\end{equation}}
\newcommand{\bem}{\begin{multline}}

\begin{document}


\title{Singularities of Floquet Scattering and Tunneling}

\author{H. Landa$^{1,2}$}
\email{e-mail: haggaila@gmail.com}
\affiliation{$^1$ LPTMS, CNRS, Univ.~Paris-Sud, Universit\'{e} Paris-Saclay, 91405 Orsay, France\\ $^2$ Institut de Physique Th\'{e}orique, Universit\'{e} Paris-Saclay, CEA, CNRS, 91191 Gif-sur-Yvette, France}

\begin{abstract}

We study quasi-bound states and scattering with short range potentials in three dimensions, subject to an axial periodic driving. We find that poles of the scattering S-matrix can cross the real energy axis as a function of the drive amplitude, making the S-matrix nonanalytic at a singular point. For the corresponding quasi-bound states that can tunnel out of (or get captured within) a potential well, this results in a discontinuous jump in both the angular momentum and energy of emitted (absorbed) waves. We also analyze elastic and inelastic scattering of slow particles in the time dependent potential. For a drive amplitude at the singular point, there is a total absorption of incoming low energy (s-wave) particles and their conversion to high energy outgoing (mostly p-) waves. We examine the relation of such Floquet singularities, lacking in an effective time independent approximation, with well known ``spectral singularities'' (or ``exceptional points''). 
These results are based on an analytic approach for obtaining eigensolutions of time-dependent periodic Hamiltonians with mixed cylindrical and spherical symmetry, and apply broadly to particles interacting via power law forces and subject to periodic fields, e.g.~co-trapped ions and atoms.

%We analyze scattering and quasi-bound states of short range potentials in three dimensions subject to an axial periodic driving. We find that poles of the scattering S-matrix can cross the real energy axis at a singular point, making the S-matrix nonanalytic, and changing abruptly the corresponding quasi-bound state. We study capture and emission processes by tunneling, elastic and inelastic scattering in the time dependent potential, and the relation of such Floquet singularities, lacking in a time independent approximation, with well known `spectral singularities' (or `exceptional points'). Being based on an analytic approach for obtaining eigensolutions of time-dependent periodic Hamiltonians with mixed cylindrical and spherical symmetry, these results apply broadly to particles subject to periodic fields and interacting via atomic power law forces.

\end{abstract}


\maketitle

\section{Introduction and main results}\label{Sec:Intro}

The main object of this paper is a time-dependent Schr\"{o}dinger equation in three dimensions, that can be brought to the form
\be i\dot{\phi }\left(\vec{r},t\right)=\left[-\frac{1}{2} \nabla ^{2} +V_{{\rm in}} \left(\vec{r},t\right)+V_{{\rm out}} \left(\vec{r},t\right)\right]\phi \left(\vec{r},t\right)\label{iphidotLR},\ee
where each potential term is dominant in a different spatial region, and both are $\pi$-periodic in time (in rescaled units in which the fundamental angular frequency is $2$, and $\hbar=m=1$).
We present an approach for obtaining approximate quasi-periodic, Floquet eigenfunctions of \eq{iphidotLR}, starting with explicitly known families of solutions for each separate Schr\"{o}dinger equation with $V_{{\rm in}}$ or $V_{{\rm out}}$, one possibly being time-independent as a particular case. This method allows us to explore a regime of parameters inaccessible to perturbation methods.

In particular we study solutions to a problem that can be formulated in two equivalent ways; one is given by the equation
\be i \dot{\psi } \left(\vec{R} ,t\right)=\left[-\frac{1 }{2 } \nabla ^{2} +V_{{\rm int}} \left(\left|\vec{R} -\vec{R}^{\pi } \left(t\right)\right|\right)\right]\psi \left(\vec{R} ,t\right)\label{ipsidot},\ee
where $V_{\rm int}$ is a spherically symmetric interaction potential and $\vec{R}^{\pi }(t)$ is a prescribed $\pi$-periodic trajectory of the center of force. To obtain a form compatible with \eq{iphidotLR}, we apply the (unitary) change of coordinates
 \be\vec{r} =\vec{R} -\vec{R}^{\pi } \left(t\right), \qquad\partial _{t} \to -\dot{\vec{R}}^{\pi } \left(t\right)\cdot \vec{\nabla }+\partial _{t}\label{rchange},\ee
and then a second unitary transformation
\be \psi =\exp \left\{i \dot{\vec{R}}^{\pi } \left(t\right)\cdot \vec{r} \right\}\phi\label{psichange},\ee
reducing \eq{ipsidot} to the sum of a time-independent central force and an additional $\pi$-periodic linear force,
\be i\dot{\phi } \left(\vec{r} ,t\right)=\left[-\frac{1}{2} \nabla ^{2} +V_{{\rm int}}\left(r\right)-\ddot{\vec{F}}\left(t\right)\cdot\vec{r}+V_F(t) \right]\phi \left(\vec{r} ,t\right)\label{iphidot0},\ee
where $r\equiv\left|\vec{r} \right|$ and
\be \vec{F}\left(t\right)=-\vec{R}^{\pi}\left(t\right),\qquad V_{F} \left(t\right)=-\frac{1}{2}\dot{\vec{R}}^{\pi } \left(t \right)^{2}.\label{Vforce}\ee
Here, the choice of what constitutes $V_{\rm in}$ and $V_{\rm out}$ depends on the approximation that is required in order to obtain the solution. The term $V_F(t)$ comes from the change of frame starting from \eq{ipsidot}, and if the starting point is \eq{iphidot0}, it will be absent. Both of these aspects will be further discussed in \seq{Sec:Numerics}, and here we keep the discussion general.

If we consider $-\ddot{\vec{F}}\left(t\right)$ to be a monochromatic electric field amplitude, and $V_{{\rm int}}\left(r\right)$  the Coulomb potential for an electron with coordinate $\vec{r}$, \eq{iphidot0} with $V_F=0$ describes the well studied problem of an atom in an AC field (the AC-Stark effect), written in the length-gauge within the dipole approximation. Then the bound states of $V_{\rm int}$ are known to turn into resonances. These are quasi-bound states with a finite lifetime determined by the imaginary part of their complex energy. This happens generally under the effect of a periodic perturbation, for any Hamiltonian with a continuous spectrum of scattering states \cite{Yajima1982,Yajima1983,Yafaev1991}. The reason is that the periodic perturbation makes every bound state with energy $\left(-\left|\epsilon\right|\right)$ resonant with unbound states from the continuum, under absorption of at least $N$ quanta from the external drive (whose frequency is $2$), where \be\left(-\left|\epsilon\right|\right)+2N >0\label{epsilon2n},\ee
and $2N$ gives the exponent of the power-law dependence of the resonance width on the perturbation amplitude.

Studies of nonperturbative violations of this picture go back to Keldysh theory \cite{Keldysh1965} and the extensive intense-laser literature \cite{0034-4885-60-4-001}. The limit where the frequency and intensity of the oscillating field are much higher than the atomic potential can be studied by using an effective averaged potential, known as the Kramers-Hanneberger (KH) approximation. This approach has led to the prediction of the remarkable phenomenon of stabilization of the atom against ionization \cite{PhysRevLett.52.613,PhysRevA.37.4536,gavrila2002atomic}, with renewed interest in recent years following experimental results \cite{eichmann2009acceleration,eichmann2013observing,0953-4075-47-20-204014} and theoretical investigations \cite{PhysRevLett.110.253001,PhysRevA.90.023401}.
 Other recent works have also revisited the systematic expansion of an effective time-independent Hamiltonian in the high-frequency limit in a general setup \cite{PhysRevLett.91.110404,PhysRevA.68.013820}, and effects related to the potential's initial phase \cite{PhysRevA.76.013421}.
For intermediate laser intensities and frequencies, the problem is inherently difficult and most approaches are based on numerical integration in some form, e.g. using close-coupled equations \cite{PhysRevLett.59.872,PhysRevA.44.R5343,PhysRevA.36.5178,PhysRevA.43.1512}, or Floquet R-matrix theory, dividing space into two regions and connecting the numerically intergated solutions at the boundary \cite{burke1991r,PhysRevA.86.043408}. 
There is renewed interest in the modeling and measuring of AC Stark shifts of trapped atoms \cite{markert2010ac}, in calculating and directly probing the angular distribution of photoelectron spectra \cite{PhysRevA.78.043403,Morales11102011}, and in the momentum distribution of emitted electrons \cite{delone1991energy,rudenko2004resonant,rudenko2005coulomb,liu2012low,xia2015momentum,ivanov2016transverse}, where cusps in the transverse momentum distribution curves were attributed to the long range nature of the Coulomb interaction.

\begin{figure}[ht]
\center {\includegraphics[width=3.1in]{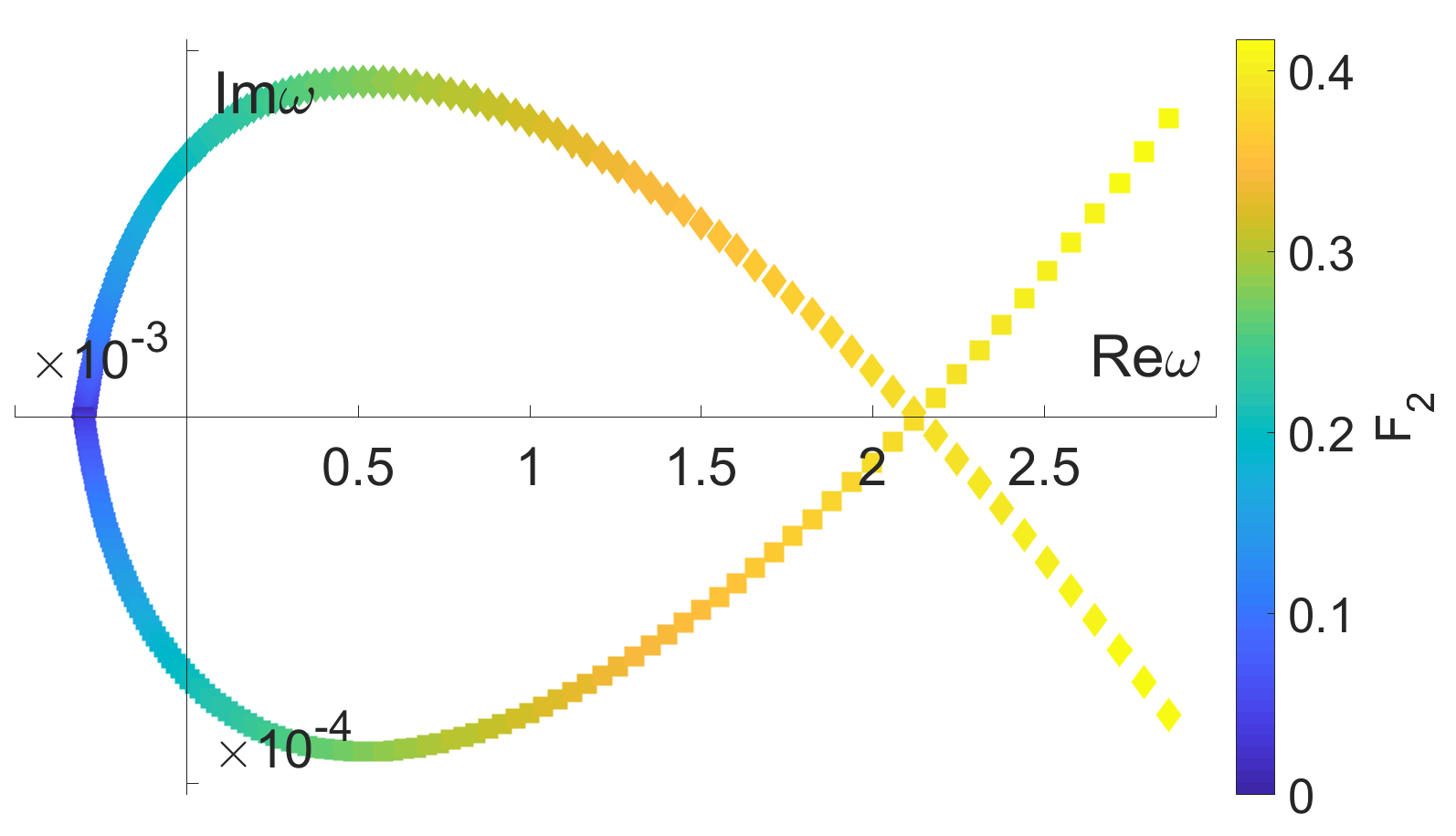}
\caption
{The complex quasi-energy $\omega$ (in nondimensional units) of two quasi-bound solutions to \eq{ipsidot} with a spherically symmetric square-well potential and an axial periodic drive (see text for details), continued with small increments of the drive amplitude $F_2$ from a near-threshold s-wave bound state. The two complex-conjugate (time-reversed) resonances are poles of the S-matrix that correspond for Im$\omega<0$ (Im$\omega>0$) to an escape (capture) process. Perturbatively in $F_2$, the escape pole is a superposition of bound components (channels of energy $\omega+2j$ with $j\leq 0$), and outgoing waves ($j\ge 1$, predominantly a p-wave with energy $(2+ \omega)\gg \omega$).
At the crossing of Im$\omega=0$ the two poles lie on different edges of the branch cut of energy plane, the scattering amplitude becomes nonanalytic at a real energy and the solution characteristics change abruptly. This is partly similar to a `spectral singularity' (or an `exceptional point') -- however unitarity is obeyed and such a solution cannot be obtained from a time-independent effective potential.
After crossing, the pole coming from above is now the emitting solution with Im$\omega < 0$, but radiates only in $j = 0$ channels (mostly s-wave of energy $\omega$) -- i.e.~the radiation is emitted by tunneling without any quanta being absorbed from the drive. In a scattering experiment, the singular point corresponds to total absorption of s-waves at the critical energy and their conversion into (mostly) higher energy p-waves.
\label{Fig:OmegaSpace0}}}
\end{figure}
%, and as it passes Re$\omega>0$, the $j=0$ channels open, but consist of asymptotically decaying incoming waves -- capturing back-reflected (mostly s-) waves.
%whose $j \ge 1$ channels are all incoming and decaying (back-reflected) waves, and it 

In contrast, in this work we focus on the case of a short range potential $V_{\rm int}$, for which Floquet resonances are much less studied. At the same time, the singularities that result from the periodic driving can be clearly identified, avoiding the additional complexity related to the Coulomb force and the accumulation of bound states near the threshold. Indeed some detailed studies of resonances in periodically driven problems were restricted to simpler one-dimensional (1D) models \cite{PhysRevA.37.98,PhysRevA.40.5614,grossmann1991coherent,PhysRevA.45.6735,moiseyev1991multiphoton,ben1993creation,timberlake2001phase,emmanouilidou2002floquet,PhysRevA.69.062105,PhysRevA.71.012102,PhysRevA.85.023407}, and include the appearance and annihilation of bound states in the dressed potential, resonant coupling between internal levels, and coherent destruction of tunneling. 

In order to study a truly 3D geometry \cite{0953-4075-35-13-302,KRAJEWSKA20082639}, in \seq{Sec:Matching} we present the main tool employed in this work, an expansion for problems with mixed cylindrical and spherical symmetry, as in \eq{iphidot0}. In \seq{Sec:Numerics} we apply this expansion to $V_{\rm int}$ that is a spherically symmetric square-well potential [\eq{squarewell}], with an additional axial force, directed along the $z$ axis, harmonic in time, 
\be \vec{F}= F_2\cos(2t)\hat{z}.\label{eq:F0}\ee
 The study of the time-{independent} square-well potential \cite{nussenzveig1959poles} constitutes one of the few examples of a detailed enumeration of the poles of the S-matrix, and their evolution in complex momentum space as a function of the universal parameter of the problem. The S-matrix is the operator that relates incoming spherical waves to outgoing, regarded as a function of complex energy or momentum \cite{landau1981quantum}, and whose poles give the bound or quasi-bound states of the system, discussed more in \seq{Sec:Quasibound}. A systematic study of the evolution of resonances subject to a periodic drive would allow a deeper understanding of the nonperturbative regime up to the high-frequency stabilization limit, and we take here the first step in this direction.

Figure~\ref{Fig:OmegaSpace0} depicts the  scenario standing at the center of this work (with the specific parameters given in \eq{Eq:Fig1params} in \seq{Sec:Approximation}). By the Bloch-Floquet theorem, solutions of \eq{iphidotLR} can always be written as a superposition of quasi-periodic wavefunctions of the Floquet form
\be \phi\left(\vec{r},t\right)=e^{-i\omega t}\phi^{\pi}\left(\vec{r},t\right)\label{Floquetansatz}\ee
where $\phi^{\pi}\left(\vec{r},t\right)$ is $\pi$-periodic. Wavefunctions of the form of \eq{Floquetansatz} constitute the equivalent of the eigenfunctions of a time-independent Hamiltonian, being characterized by a single frequency $\omega$, the (quasi-) energy. 
Hence, in  \fig{Fig:OmegaSpace0}, two exact quasi-periodic solutions are followed in complex $\omega$-plane by continuation as $F_2$ is increased. These solutions give poles of the S-matrix, as defined by the boundary conditions.  For $F_2=0$ both coincide at an s-wave bound state of the time-independent square well, and thus lie initially within the physical sheet of complex energy. Since we consider a time-reversal invariant Hamiltonian, the two poles are related by complex conjugation in $\omega$ plane.

A quasi-bound state with a general complex $\omega$ is a coherent superposition of components bound to the potential well, and components which are asymptotically (for $r\to \infty$) travelling waves (going inwards or outwards). As we discuss in \seq{Sec:Relations}, since \eq{ipsidot} becomes in this limit the equation of a free particle, the solutions (in that frame) tend to a superposition of free spherical waves (in the frame of \eq{iphidot0} the waves remain periodically driven). According to \eq{Floquetansatz} each such component must have an energy equal to $\omega+2j$ with $j\in\mathbb{Z}$. 
The form of a wavefunction of a complex (quasi-) energy can be understood in the limit of $F_2\ll 1$ using perturbation theory as mentioned above.
For the pole with Im$\omega<0$ the probability of measuring the state in one of the localized components decreases with time and hence there must be a corresponding escaping probability flux. The open channels (all partial waves with energy Re$\omega+2j>0$) are therefore  outgoing waves, diverging at $r\to\infty$. The corresponding momentum $k_{2j}$, defined by
\be\frac{1}{2}k_{2j}^2=\omega+2j,\label{eq:k2j}\ee
must have Re$k_{2j}>0$ and Im$k_{2j}<0$ for $j\ge 1$ at $F_2\to 0$.
 For the time-reversed solution pole, $\omega$ goes into the upper half plane as $F_2$ is increased from 0, the probability increases with time and the open channels must correspond to (diverging) incoming waves (with the root of \eq{eq:k2j} chosen to have Re$k_{2j}<0$ and Im$k_{2j}<0$) -- this is a capture process. The closed (bound) channels are exponentially decaying in both cases with Im$k_{2j}>0$. 
This choice of boundary conditions is then held fixed and defines the solutions that are continuously followed as $F_2$ varies. Each channel is followed separately and continuously -- each $k_{2j}$ moves on its own Riemann surface \cite{newton2013scattering,PhysRevA.61.032716} as the parameter of the continuation is varied.

In contrast to free waves with real momentum, that can be obtained as the limit of square integrable eigenfunctions of the free particle Hamiltonian, the diverging (Gamow-Siegert \cite{rosas2008primer}) waves with complex momentum are eigenfunctions of an (asymptotically) non-Hermitian free particle Hamiltonian. Non-Hermitian Hamiltonians \cite{rotter2009non} very generally are the result of tracing out some degrees of freedom, and the resulting non-unitarity is a consequence of probability flux going into those degrees of freedom, that now lie outside the Hilbert space.
The diverging waves carry a well defined probability current density, and thus can be used to calculate a relative probability flux.
If a quasi-bound state is decaying, the relative probability to detect an outgoing wave in one of the open channels will be proportional to the square of its amplitude (see \seq{Sec:Quasibound} and \fig{Fig:Rate1}). In the time reversed process,  the bound state can form if particles are sent in towards the center, with the probability for this process again proportional to the overlap of available free particle states with the Floquet solution, and (half) the rate of formation is given by Im$\omega$.

%If the decay rate is small enough (and smaller than level separation), an experiment starting with the undriven bound state and adiabatically increasing the drive strength (but faster than the decay rate) will allow the solution to `follow' the pole, with some finite probability to remain in the bound state. 

 Alternatively, a standard scattering formulation (discussed in \seq{Sec:Scattering}) is obtained by  setting $\phi\left(\vec{r},t\right)$ to be asymptotically the sum of a plane wave and scattered outgoing spherical waves, and restricting $\omega$ to be the real, positive energy of the plane wave. Resonances in the scattering cross section can be typically related to quasi-bound (resonance) states \cite{newton2013scattering}, and hence the nomenclature coincidence. The relation results from the scattering amplitude (and the S-matrix) being a meromorphic function of the energy (or momentum) in some region in complex plane (that depends on the potential), and hence its values on the real axis are significantly influenced by nearby poles. We follow both aspects and their relation in our analysis of singularities in the Floquet setup.

Indeed, as $F_2$ is increased, the absolute value of ${\rm Im}\omega$ in \fig{Fig:OmegaSpace0} initially increases and then decreases -- this is an example of a nonperturbative stabilization at higher field amplitudes. As the pole in the lower half-plane passes to Re$\omega>0$, the $j=0$ channels open, but  consist of asymptotically decaying incoming waves (in momentum space, $k_0$ just crosses the bisector of its quarter-plane) -- capturing back-reflected (mostly s-) waves. 
At a critical value of the drive strength the two poles reach the real $\omega$-axis and at this singular point with a real energy, the nature of the quasi-bound states changes abruptly. We  denote the parameters of this point by
\be F_2=\bar{F}^c, \qquad \omega = \bar{\omega}^c.\label{Fomegac}\ee
Just before crossing the real line, the solution with Im$\omega<0$ radiates partial waves of energy $\omega+2j$ with $j\ge 1$ after absorbing at least one quantum from the drive, with the dominant partial wave being the p-wave with $j=1$. After crossing, this solution becomes the capture solution (with Im$\omega>0$) and all previously open radiating channels are now decaying (Im$k_{2j}>0$ for $j\ge 1$) and carry no current asymptotically. This solution is capturing incoming (mostly s-) waves in the $j=0$ channels (which have crossed to Im$k_{0}<0$, while remaining with Re$k_{0}<0$). The radiating solution now, for $F_2>\bar{F}^c$, is the solution that came from the upper (energy) half plane. All channels with $j\ge 1$ (which were incoming, diverging waves) are now asymptotically decaying -- in the frame of \eq{iphidot0} they are back reflected by the oscillating drive  (in the frame of \eq{ipsidot} it is destructive interference of waves emitted during the oscillations). The only open, radiating channels are with $j=0$ (mostly the s-wave), which was previously asymptotically decaying and are now diverging. Thus the bound state radiates solely by tunneling, without absorbing quanta from the drive.

Exactly at Im$\omega=0$ (for $\bar{F}^c$) there is one pole on the upper edge of the cut of complex energy plane (this is the pole that moved through the upper half of the plane), and one pole directly below it, on the lower edge of the cut. Both are about to leave the physical sheet.
In momentum space, the two corresponding $k_{0}$ poles are crossing from the upper to the lower half plane, on both sides of the imaginary axis. The two solutions  have a real (degenerate) energy and the $j\ge 0$ channels are (driven) free particle waves. The solutions describe a balanced flux of incoming and outgoing waves in the respective channels. This is a `self-sustaining' standing wave, that exists with open boundary conditions.
The S-matrix becomes nonanalytic on the real energy axis for this critical parameter, and an effective, time-averaged Hermitian approximation of the potential (as in the KH approximation discussed above) cannot result in such a solution, as this would violate unitarity of the resulting elastic scattering.

In the full time-dependent setup however, the scattering is inelastic and unitarity is obviously not violated. In terms of the S-matrix, paired with each pole there is a zero of the S-matrix that in momentum space is located at $-k_{2j}$, which follows the same trajectory in energy plane. In a scattering experiment at  $F_2=\bar{F}^c$ where the poles and zeros all coincide at $\omega=\bar{\omega}^c$, an incoming ($j=0$) s-wave with this value of energy is completely absorbed and removed from the scattered wavefunction -- which becomes predominantly p-wave with $j=1$. This occurs as the scattering amplitude of the incoming s-wave in the $j=0$ channel goes through $0$ at the singular point, and in a 2D parameter space composed of $F_2$ and the real scattering energy $\omega$, a $2\pi$ phase is accumulated around this point (see \fig{Fig:s-waveimages}).

Hence the time-dependent solution at the critical point of crossing the real $\omega$-axis shares some properties with singularities discussed mostly in the context of time-independent complex potentials. These include in  particular `spectral singularities', that occur with two scattering states with a real energy in a complex potential \cite{mostafazadeh2015physics}. However with spectral singularities the manifestly complex potential violates unitarity, which is not the case in the current setup.
Similarly, an `exceptional point' typically refers to the coalescence of two discrete states of a complex Hamiltonian \cite{heiss2012physics}, a problem that continues to be studied theoretically \cite{gilary2012asymmetric,gilary2013time,milburn2015general}, with interesting recent realizations and implications \cite{zhen2015spawning,doppler2016dynamically,xu2016topological}. In contrast, in the current problem, despite the coincidence of two poles at the same (real) value of $\omega$, there is no coalescence of the eigenvectors. The coincidence of a pole and a zero in the Floquet problem is also similar to singular points of laser-absorber $\mathcal{P}\mathcal{T}$-symmetric systems \cite{PhysRevLett.105.053901,PhysRevA.82.031801,PhysRevLett.106.093902}, which are again nonunitary. In such optical systems with gain and loss also Floquet setups attract increasing interest \cite{PhysRevLett.119.093901,PhysRevA.96.042101}. Exceptional points in a Floquet unitary scattering setup have been discussed before \cite{magunov2001laser,atabek2011proposal}, and we further discuss Floquet exceptional points in \seq{Sec:Outlook}. We note also that the solution at the critical point is not a `bound state in the continuum' \cite{newton2013scattering,mostafazadeh2012spectral}, since it is not square-integrable.

Further relevant examples where the results presented here are applicable include quantum wires and dots \cite{leyronas2001quantum} that have been modeled by similar finite-barrier potentials, and the expansion presented here can be used to solve a mixed-type system, periodically driven.
Interacting cold atoms or molecules \cite{carr2009cold}  are often subject to oscillating fields \cite{PhysRevLett.94.083001,tokunaga2011prospects}. The generalization to settings with a potential of spherical symmetry in the exterior region is straightforward, and the case of zero-range interaction has been recently treated in  \cite{ZeroFloquet}. Overlapping a trap for neutral atoms with a periodically driven Paul trap for ions \cite{WinelandReview} was suggested and realized \cite{PhysRevA.67.042705,PhysRevLett.102.223201}, followed by the demonstration of a trapped ion immersed in a dilute atomic Bose-Einstein condensate \cite{zipkes2010trapped,PhysRevLett.105.133202}, and many other experiments. The effect of the periodic drive of the ion has been analyzed for classical collisions with the atom \cite{PhysRevLett.109.253201} or Rydberg atoms \cite{PhysRevLett.118.263201}, for quantum scattering employing a master equation description \cite{PhysRevA.91.023430}, and for an ion and atom in separate traps \cite{PhysRevA.85.052718}.
Quantum Defect Theory (QDT) \cite{PhysRevA.26.2441,Seaton1983} is a very important theoretical tool for modeling  atomic scattering, and continues to evolve \cite{PhysRevA.78.012702,PhysRevA.79.010702,PhysRevA.75.053601,PhysRevA.80.012702,PhysRevLett.104.213201,PhysRevA.84.042703,idziaszek2011multichannel,PhysRevA.88.022701,PhysRevLett.110.213202,PhysRevA.87.032706}, together with new models and methods \cite{PhysRevA.73.063619,PhysRevA.82.042712}, applied to many-body states as well \cite{PhysRevA.90.033601,schurer2015capture,PhysRevLett.119.063001}. As we argue in \seq{Sec:Outlook}, the results presented in this work hold for short range power-law potentials, and calculations using QDT, that can naturally be used in the interior region, show that they can potentially be observed with a co-trapped ion and atom system \cite{unpublished}.

This paper is organized as follows. \seq{Sec:Matching} develops the theory. In \seq{Sec:Cylindrical} we introduce the expansion that is used in \seq{Sec:Matching3D} to relate the solution matching conditions. Some general properties of the solutions are discussed in \seq{Sec:Relations}, and then \seq{Sec:Quasibound} discusses in more detail quasi-bound (pole) solutions and their characterization, while \seq{Sec:Scattering} introduces scattering solutions and cross sections. We conclude the formalism with a review of some analytic properties of the wavefunctions used in the expansion, and of the partial waves expansion, in \seq{Sec:Analytic}. The results of applying the theory to the problem of a driven square well are presented in \seq{Sec:Numerics}, with a driven loosely bound s-wave studied in \seq{Sec:Looselys-wave} and a deeper bound p-wave in \seq{Sec:p-wave}. In \seq{Sec:Approximation} we discuss some aspects of the method and compare approximations that can be achieved with it, concluding in \seq{Sec:Outlook} with a discussion of the applicability of this work to more physical atomic potentials, and the relation to other singularities of non-Hermitian Hamiltonians. The Appendices contain, in addition to some details of the derivation, a few general expressions useful for the calculation of expectation values using the solution wavefunctions.

%, e.g. by all wavefunctions of a specific $\pi$-periodic Hamiltonian. Then, any other $\pi$-periodic Hamiltonian can be expanded using such a basis for the extended Hilbert space, and all of the tools of time-independent quantum theory are available, which can be powerful in many scenarios, e.g. for employing perturbation theory. 

\section{Floquet wavefunctions}\label{Sec:Matching}

%In this section we formulate a method for finding wavefunctions of \eq{iphidotLR}, with a time-independent potential $V_{{\rm in}} \left(\vec{r}\right)$ which is assumed to be significant inside some interior region $\left|\vec{r}\right|<d$, and a time-dependent $\pi$-periodic potential $V_{{\rm out}} \left(\vec{r},t\right)$ which dominates in the exterior region $\left|\vec{r}\right|>d$. The  assumptions at the basis of the presented approach are that the wavefunctions of each of the potentials can be found explicitly, and that there is some meaning to dividing space into the interior and exterior regions, even if only as a (zeroth-order) approximation.

%In the asymptotic region $R\to\infty$ (with $R=\left|\vec{R}\right|$), the solution of \eq{ipsidot} is given by \be \phi_{\rm sc}\left(\vec{R},t\right)\sim \sum_{j,l_1}b_{j,l_1}e^{-i(\omega+2j)t} h_{l_1}^{\left(a\right)} \left(k_{2j}R\right) Y_{l_1}^{m} \left(\theta,\varphi\right),\label{psiRasymp}\ee with  \be \frac{1}{2}k_{2j}^2= \omega+2j.\ee We denote the above wavefunction by $\phi_{\rm sc}$ because in a scattering setup it is just the scattered part of the solution. 

\subsection{Floquet waves with cylindrical symmetry}\label{Sec:Cylindrical}

Starting with the general time-dependent Schr\"{o}dinger equation 
\be i \dot{\phi }\left(\vec{r},t\right)=\left[-\frac{1 }{2} \nabla ^{2} -\ddot{\vec{F}}\left(t\right)\cdot \vec{r} +V_{1} \left(t\right)\right]\phi \left(\vec{r},t\right)\label{ihbarphidot3Dext},\ee
 a family of solutions can be written in the form
\be \phi \left(\vec{r},t\right)\propto e^{ i\vec{q}\left(t\right)\cdot \vec{r}-ig\left(t\right)},\qquad \vec{q}\left(t\right)=\dot{\vec{F}} +\vec{k},\ee where $\vec{k}$ is the (possibly complex) constant of integration, and 
\be {g\left(t\right)=\frac{1 }{2} \vec{k}^{2} t}\\+\int _{}^{t}\left[ \vec{k}\cdot \dot{\vec{F}}\left(t'\right)+\frac{1}{2}\dot{\vec{F}}\left(t'\right)^{2} +V_{1} \left(t'\right) \right]dt'.\label{gtint}\ee

In the rest of this paper, we assume the drive to be $\pi$-periodic and coaxial at any time, and choose a cylindrical coordinate system, $\vec{r}=\left(\rho,z,\varphi\right)$, in which 
\be {\vec{F}}\left(t\right)={F}^{\pi } \left(t\right)\hat{z}\label{Fpi}.\ee
Furthermore, if \eq{ipsidot} is the physical starting point we have using \eq{Vforce}
\be V_1\left(t\right)=V_F\left(t\right)=-\frac{1}{2}\dot{\vec{F}}\left(t\right)^{2},\label{V1}\ee
which also simplifies the current expressions by  cancelling the $\vec{k}$-independent term in \eq{gtint}. We return to this point in \seq{Sec:Approximation}.

The cylindrical waves are eigenfunctions of the free particle Hamiltonian, $H=-\frac{1 }{2} \nabla ^{2}$, given by
\be \chi_{m}^{\left(1,2,J\right)} \left(\vec{r};k,\alpha \right)=  e^{im\varphi }H_{m}^{\left(1,2,J\right)} \left(k\rho \sin \alpha \right)e^{ikz\cos \alpha },\label{chim12J}\ee
where $m$ is the magnetic quantum number, $k$ is the (possibly complex) wavenumber, and $\alpha$ is a complex parameter. $H_{m}^{\left(1,2,J\right)}$ is a Hankel function of the first or second kind (corresponding to outgoing and incoming traveling-waves respectively), or a Bessel function (which we denote with a superscript $J$).
Thus outgoing and incoming traveling-wave solutions to \eq{ihbarphidot3Dext}, subject to \eqs{Fpi}-\eqref{V1}, can be written using \eq{chim12J} in the form
\bem \phi _{m}^{\left(1,2\right)} \left(\vec{r},t;k,\alpha \right)\\ = e^{-i\frac{1}{2} k^{2} t} e^{i\dot{F}^{\pi } \left(t\right)z} \chi _{m}^{\left(1,2\right)} \left(\vec{r};k,\alpha \right)e^{-iF^{\pi } \left(t\right)k\cos \alpha }\label{phiRm12}.\end{multline}
Equation \eqref{phiRm12} is a particular quasi-periodic solution of \eq{ihbarphidot3Dext}, taking the Floquet form of \eq{Floquetansatz}. With $k_{2j}$ defined in \eq{eq:k2j}, the most general quasi-periodic solution to \eq{ihbarphidot3Dext}  is
\be \phi _{m} \left(\vec{r},t\right)=\sum _{\substack{j\in \mathbb{Z}\\a=1,2}}\int_{C^{(a)}} d\alpha \sin\alpha b_{2j}^{\left(a\right)} \left(\alpha \right)\phi _{m}^{\left(a\right)} \left(\vec{r},t;k_{2j} ,\alpha \right),\label{phimexterior}\ee
which consists at each value of $k_{2j}$ of a  superposition of outgoing and incoming cylindrical waves, parametrized by integrals in complex $\alpha$-plane along the contours $C^{(a)}$ with weight functions $b_{{2j}}^{\left(a\right)}\left(\alpha\right)$.

%Bessel functions with $k>0$ and real $\alpha$ have the momentum space normalization \bem \int d\varphi dz d\rho \rho \left[\chi_{m'}^{\left(J\right)} \left(\vec{r};k',\alpha' \right)\right]^* \chi_{m}^{\left(J\right)} \left(\vec{r};k,\alpha \right)\\=({2\pi})^2\delta_{mm'}\delta(k_\parallel -k'_\parallel)\delta(k_\perp -k'_\perp)/k_\perp,\end{multline} where $k_\parallel=k\cos\alpha$ and $k_\perp=k\sin\alpha$.

To transform the solution of \eq{phimexterior} to spherical coordinates $\vec{r}=\left(r,\theta,\varphi\right)$, we take the arbitrary weight functions for cylindrical Floquet waves in \eq{phimexterior} to be
\be b_{{2j} }^{\left(a\right)} \left(\alpha \right)=\sum _{l_{1} }b_{2j,l_{1} }N_{2j,l_1}^m  S_{2j,l_1}^{\left(a\right)} P_{l_{1} } \left(\cos \alpha \right),\label{be2jm} \ee
 with $N_{2j,l_1}^m$ a normalization constant to be defined in \eq{scatteringnormalization}, $S_{2j,l_1}^{\left(a\right)}$ will depend on the boundary conditions at $r\to\infty$, and $b_{2j,l_{1} }$ will be matching coefficients for quasi-bound states, as elaborated in the following subsections. In \app{Sec:Derivations1} we show that by plugging \eq{be2jm} into \eq{phimexterior}, each term of the series within the latter equation can be rewritten as
\bem {\int_{C^{(a)}} d\alpha \sin\alpha b_{{2j} }^{\left(a\right)} \left(\alpha \right)\phi _{m}^{\left(a\right)} \left(\vec{r},t;k_{2j} ,\alpha \right)}=\\ {e^{-i\frac{1}{2} k_{2j} ^{2} t}\sum _{l_{1},l }b_{2j,l_{1} }N_{2j,l_1}^m S_{2j,l_1}^{\left(a\right)} R_{2j,l,l_{1} } ^{\left(a\right)}\left(r,t\right) Y_{l}^{m} \left(\theta ,\varphi \right), }\label{intphiRexpansion} \end{multline}
%\be \begin{array}{l} {\int_{C_1} d\alpha \sin\alpha b_{{2j} }^{\left(1\right)} \left(\alpha \right)\phi _{R,m}^{\left(1\right)} \left(\vec{r},t;k_{2j} ,\alpha \right)}\\\\{=e^{-i\frac{1}{2} k_{2j} ^{2} t} e^{i\dot{F}^{\pi } \left(t\right)z}\times }\\\\{\quad\int_{C_1} d\alpha \sin \alpha b_{2j }^{\left(1\right)}\left(\alpha \right)\chi _{m}^{\left(1\right)} \left(\vec{r};k_{2j} ,\alpha \right)e^{-iF^{\pi } \left(t\right)k_{2j} \cos \alpha }  } \\\\ {=e^{-i\frac{1}{2} k_{2j} ^{2} t}\sum _{l_{1},l }b_{l_{1} ,2j } R_{l,l_{1} ,2j } \left(r,t\right) Y_{l}^{m} \left(\theta ,\varphi \right)  } \end{array}\ee
where $Y_{l}^m$ are normalized spherical harmonics [\eq{Nlmdefinition}] and the radial functions are 
\bem R_{2j,l_{1} ,l}^{\left(a\right)} \left(r,t\right)=\\\sum _{l_{2} ,l_{3} ,l_{4} }c_{l_{1} ,l_{2} ,l_{3} ,l_{4} ,l} j_{l_{2} } \left(F^{\pi } \left(t\right)k_{2j} \right)j_{l_{4} } (\dot{F}^{\pi } \left(t\right)r) h_{l_{3} }^{\left(a\right)} \left(k_{2j} r\right),\label{R2jl1la}\end{multline}
with the coefficients $c_{l_{1} ,l_{2} ,l_{3} ,l_{4} ,l} $ being defined in \eq{cl1l2l3l4l}, and the spherical Hankel functions of the first (second) kind, $h_{l}^{\left(1\right)}$ ($h_{l}^{\left(2\right)}$), correspond to outgoing (incoming) spherical waves (for ${\rm Re}k>0$). This expansion forms the essential tool that allows, in conjunction with the matching described in the following subsection, to obtain the results in this work.

\begin{widetext}

\subsection {Floquet wavefunctions with mixed cylindrical and spherical symmetry}\label{Sec:Matching3D}

Adding a spherically symmetric interaction potential to \eq{ihbarphidot3Dext} [with $V_1$ of \eq{V1}], we regain \eq{iphidot0},
\be i \dot{\phi }\left(\vec{r},t\right)=\left[-\frac{1 }{2} \nabla ^{2} +V_{{\rm int}}\left(r\right)-\ddot{\vec{F}}\left(t\right)\cdot \vec{r} +V_{F} \left(t\right)\right]\phi \left(\vec{r},t\right).\label{iphidot3D}\ee
Depending on $V_{\rm int}$, this equation may be solvable either exactly or only approximately. The solution proceeds by assuming that \eq{iphidot3D} can be replaced by an equation having the form of \eq{iphidotLR},
\be i\dot{\phi }\left(\vec{r},t\right)=\left[-\frac{1}{2} \nabla ^{2} +V_{{\rm in}}^{(1)} \left(\vec{r},t\right)+V_{{\rm out}}^{(1)} \left(\vec{r},t\right)\right]\phi \left(\vec{r},t\right)\label{iphidotLR2},\ee
and dividing space into two regions, interior and exterior to sphere $\left|\vec{r}\right|=d$, where the Schr\"{o}dinger equation can be solved exactly with either one of the potentials above. In this section we will focus on the case when an approximation is required, taking the form
\be V_{{\rm in}}^{(1)} \left(\vec{r},t\right)=V_{{\rm int}}\left(r\right)\Theta(d-r),\qquad V_{{\rm out}}^{(1)} \left(\vec{r},t\right)=\left[-\ddot{\vec{F}}\left(t\right)\cdot \vec{r} +V_{F} \left(t\right)\right]\Theta(r-d) \label{Vin1Vout1},\ee
where $\Theta(\cdot)$ is the Heaviside function. Thus in \eq{iphidot3D} we have truncated $V_{\rm int}$ at a finite radius and removed the external drive from the interior region, leaving it to modulate the free-particle exterior region. A further discussion of this approximation (and a comparison to the exact solution for a square well) will follow in \seq{Sec:Approximation}. For finding  quasi-periodic solutions in the Floquet form of \eq{Floquetansatz}, we can employ the ansatz
\be \phi _{m} \left(\vec{r},t\right)=\left\{\begin{array}{ccc} {\sum_{n,l}a_{2n,l} e^{-i\left(\omega +2n\right)t} \phi _{{\rm in},\omega +2n,l}\left(r\right) Y_{l}^m\left(\theta,\varphi\right)  } & & {r<d} \\\\ {\sum _{ j,l_{1}} b_{2j,l_{1} }e^{-i\left(\omega +2j\right)t}\sum _{l}\phi_{{\rm out},2j,l_1,l}^{\pi} \left(r,t\right)Y_{l}^m \left(\theta ,\varphi \right)} & & {r>d} \end{array}\right.\label{phi3Dansatz},\ee
with
\be \phi_{{\rm out},2j,l_1,l}^{\pi} \left(r,t\right)= N_{2j,l_1}^m \left[S_{2j,l_1}^{\left(2\right)} R_{2j,l_{1},l }^{\left(2\right)} \left(r,t\right) + S_{2j,l_1}^{\left(1\right)}R_{2j,l_{1},l }^{\left(1\right)} \left(r,t\right)\right]. \label{phi3DS2S1} \ee

This wavefunction is the most general Floquet superposition of solutions in both the interior and exterior region and is an exact solution of the original problem [\eq{iphidot3D}] in both limits $r\to 0$ and $r\to\infty$. In the interior region each wavefunction $\phi _{{\rm in},\omega +2n,l}\left(r\right)$ is (locally) a solution with energy $\omega+2n$ of the Schr\"{o}dinger equation with the potential $V_{\rm in}^{(1)}\left(\vec{r}\right)$. The Fourier expansion (that is tractable if it can be truncated of course) takes all integers $n\in \mathbb{Z}$ (and similarly for $j$),
 without any a-priori restriction to positive (real part) energies. The required boundary conditions at $r\to 0$ are assumed to have been imposed, which determine a unique linear combination of the two linearly independent solutions at each value of $\omega+2n$. This can be the condition of regularity at the origin, or a quantum defect theory parametrization. The boundary condition at $r\to \infty$ are discussed in the following subsections. Finally, \eq{phi3DS2S1} does not indicate explicitly a summation over any degeneracy of the wavefunctions, which might require independent matching coefficients.

Requiring the continuity of the wavefunction across the surface of the sphere $\left|\vec{r}\right|=d$ gives the equation (trivially independent of $\theta,\varphi$ due to the identical expansion in spherical harmonics on both sides)
\bem \sum _{n,l}e^{-i\left(\omega +2n\right)t} a_{2n,l}  \phi _{{\rm in},\omega +2n,l} \left(d\right)  =\sum _{ j,l_{1}} e^{-i\left(\omega +2j\right)t}b_{2j,l_{1} }\sum _{l} \phi_{{\rm out},2j,l_1,l}^{\pi} \left(d,t\right) \\  =\sum _{ j,l_{1}} e^{-i\left(\omega +2j\right)t}b_{2j,l_{1} }\sum _{l,p} d_{2p,l,2j,l_{1} } e^{-i2pt}     =\sum _{ n,l} e^{-i\left(\omega +2n\right)t}\sum _{j,l_1}b_{2j,l_{1} } d_{2\left(n-j\right),l,2j,l_{1} } \end{multline}
%\be \begin{array}{l} {\sum _{n,l}e^{-i\left(\omega +2n\right)t} a_{2n,l}  \phi _{L,\omega +2n,l} \left(d\right) Y_{l}^m  } \\\\ {\qquad =\sum _{ j,l_{1}} e^{-i\left(\omega +2j\right)t}b_{2j,l_{1} }\sum _{l} \phi_{R,2j,l_1,l}^{\pi} \left(d,t\right) Y_{l}^m  } \\\\ {\qquad =\sum _{ j,l_{1}} e^{-i\left(\omega +2j\right)t}b_{2j,l_{1} }\sum _{l,p} d_{2j,l_{1},l,2p } e^{-i2pt}Y_{l}^m  } \\\\ {\qquad =\sum _{ n,l} e^{-i\left(\omega +2n\right)t}\sum _{j,l_1}b_{2j,l_{1} } d_{2j,l_{1},l,2\left(n-j\right) } Y_{l}^m   } \end{array}\ee
where $d_{2p,l,2j,l_{1} } $ are the expansion coefficients of the Fourier series of $\phi_{{\rm out}	,2j,l_1,l}^{\pi} \left(d,t\right)$, which in general must be obtained numerically.
Then the first matching condition is
\be c_{2n,l}a_{2n,l}\equiv\phi _{{\rm in},\omega +2n,l} \left(d\right) a_{2n,l} =\sum _{j,l_1} d_{2\left(n-j\right),l,2j,l_{1} } b_{2j,l_{1} }.\label{match1}\ee
Similarly, the second matching condition comes from the continuity of the radial derivative ($\partial_r$), which gives
\be f_{2n,l}a_{2n,l}\equiv\partial _{r} \phi _{{\rm in},\omega +2n,l} \left(d\right)a_{2n,l} =\sum _{j,l_1} g_{2\left(n-j\right),l,2j,l_{1} } b_{2j,l_{1} },\label{match2}\ee
with $g_{2p,l ,2j,l_{1}}  $ the expansion coefficients of the Fourier series of $\partial _{r} \phi_{{\rm out},2j,l_1,l}^{\pi} \left(d,t\right)$. The latter  can be written as
\be \begin{array}{l} {\partial _{r} R_{2j,l_{1} ,l }^{\left(a\right)} \left(r,t\right)=\sum _{l_{2} ,l_{3} ,l_{4} }c_{l_{1} ,l_{2} ,l_{3} ,l_{4} ,l} j_{l_{2} } \left(F^{\pi }\left(t\right) k_{2j} \right)} \\ \\ {\qquad  \times\left\{-\dot{F}^{\pi }\left(t\right) j_{l_{4} +1} \left(\dot{F}^{\pi }\left(t\right) r\right)h_{l_{3} }^{\left(a\right)} \left(k_{2j} r\right)+j_{l_{4} } \left(\dot{F}^{\pi }\left(t\right) r\right)\left[\frac{1}{r}\left(l_{3} +l_{4} \right) h_{l_{3} }^{\left(a\right)} \left(k_{2j} r\right)-k_{2j} h_{l_{3} +1}^{\left(a\right)} \left(k_{2j} r\right)\right]\right\}.} \end{array}\ee
In the following subsections we elaborate on the properties of the wavefunction, and we will arrange these recursion formulas in a matrix form, to facilitate their solution. % That form  depends on the  boundary conditions, and in particular, on whether quasi-bound states or scattering solutions are sought. Some properties of the solutions however, are shared between both types and are elaborated in the following subsection. 
\end{widetext}

\break

\subsection{General properties of the solutions}\label{Sec:Relations}

Some general properties of the Schr\"{o}dinger equation and its solutions will be used in the following. First, we assume that $\vec{F}(t)$ [or $\vec{R}^\pi(t)$] is time-reversal invariant (including the possibility that this requires a trivial shift of $t$). Then the Schr\"{o}dinger equation is invariant under a simultaneous change $t\to-t$ and complex conjugation, and hence if $\phi(t)$ is a solution, so is $\phi^*(-t)$, which maybe the same wavefunction or an independent one.
In addition, the equation conserves probability locally in time and space, such that the continuity equation holds
\be \partial_t n(\vec{r},t) + \vec{\nabla} \cdot \vec{j}\left(\vec{r},t\right) =0,\ee
with the density and probability current density defined by
\be n(\vec{r},t) = \left|\phi\right|^2,\quad \vec{j}\left(\vec{r},t\right)=\frac{1}{2i}\left[\phi^*\nabla\phi-\phi\nabla\phi^* \right]\label{eqj}.\ee
 The continuity equation holds irrespective of the Hermiticity of the boundary conditions (or whether $\omega$ is real or complex), as long as the Hamiltonian is real. Then, {\it if} the wavefunction is square integrable in configuration space, its norm remains constant (and finite) in time. However, even if the wavefunction is not normalizable, a meaning can be attributed to the relative probability amplitude of each asymptotic channel, as detailed in the following subsection.
 
In order to gain more insight into the physical meaning of the wavefunction of \eq{phi3Dansatz}, we can use its connection to the Schr\"{o}dinger equation in the `lab' frame, 
\eq{ipsidot}, which is reproduced here again,
\be i \dot{\psi } \left(\vec{R} ,t\right)=\left[-\frac{1 }{2 } \nabla ^{2} +V_{{\rm int}} \left(\left|\vec{R} -\vec{R}^{\pi } \left(t\right)\right|\right)\right]\psi \left(\vec{R} ,t\right)\label{ipsidot2}.\ee
In the asymptotic $\vec{R}\to \infty$ region, $V_{\rm int}$ decays and the solutions of \eq{ipsidot2} reduce to free particle solutions. By starting with a free spherical wave of momentum $k_{2j}$ (dropping $e^{-i(\omega+2j) t}$ for simplicity),
\be \psi_{k_{2j}l_1m}^{(a)}\left(\vec{R}\right)=N_{2j,l_1}^m h_{l_1}^{\left(a\right)} \left(k_{2j}R\right) P_{l_1}^{m} \left(\cos\theta\right)e^{im\varphi },\label{freewaveR}\ee
effecting the unitary transformation of \eqs{rchange}-\eqref{psichange} and then using the representation of \eq{hlexpansion}, it is  seen that in fact the transformation carries
\be \psi_{k_{2j}l_1m}^{(a)}\left(\vec{R}\right) \to N_{2j,l_1}^m \sum _{l} R_{2j,l_{1},l }^{\left(a\right)} \left(r,t\right) Y_{l}^m \left(\theta ,\varphi \right),\label{freewaveRasymp}\ee
so that $b_{2j,l_1}S_{2j,l_1}^{(a)}$ in the solution \eq{phi3Dansatz} is the coefficient of the asymptotically free spherical wave in the `lab' frame (with $\vec{R}$), with energy $2j+\omega$ and angular momentum quantum number $l_1$. A similar conclusion can also be obtained in the frame of \eq{iphidot3D}. In this frame, in any physical realization, the periodic drive $\vec{F}(t)\cdot\vec{r}$ cannot continue to infinity. Then if beyond some large enough distance $r$ outside the range of $V_{\rm int}$, the amplitude of periodic drive $\vec{F}(t)$ is adiabatically diminishing (in space), then the solutions will become asymptotically free spherical waves again. However, in this frame, \eq{V1} has to be modified ($V_1=0$), and the asymptotic momenta will be different. We do not include this calculation explicitly although we return to this point in \seq{Sec:Approximation}.

The wave of \eq{freewaveR}, in the nondecaying channels, carries a momentum current density proportional to Re$k_{2j}$. Thus for the solutions of \eq{ipsidot} [\eq{ipsidot2}] we can set for the normalization constant of \eq{be2jm} and \eq{phi3DS2S1},
\be N_{2j,l_1}^m=\left({\rm Re} v_{2j}\right)^{-1/2}({\rm Re} k_{2j})N_{l_1}^m=\left({\rm Re} k_{2j}\right)^{1/2}N_{l_1}^m,\label{scatteringnormalization}\ee
with $N_l^m$ defined in \eq{Nlmdefinition} and the velocity of a particle is related to its wavenumber in nondimensional units simply by ${\rm Re}v_{2j}={\rm Re}k_{2j}$. This makes the probability current density $\vec{j}$ of \eq{eqj} in one outgoing, nondecaying channel wavefunction [\eq{freewaveR}], asymptotically normalized to unit flux on the sphere, 
\be\vec{j}\sim \left|Y_{l_1}^m \left(\theta ,\varphi \right)\right|^2\hat{r}.\label{asymptoticj}\ee
This normalization will be used in the following two subsections.

\subsection{Quasi-bound wavefunctions}\label{Sec:Quasibound}

Quasi-bound states are defined by fixing the  boundary conditions  in all channels, by setting each of the constants $S_{2j,l_1}^{\left(a\right)}$  in \eq{phi3DS2S1}, to either 0 or a modulus 1 value, $\left|S_{2j,l_1}^{\left(a\right)}\right|^2 = \left\{0,1\right\}$.
 The two matching relations can be written in matrix form (once a finite truncation has been applied),
\be C\vec{a}=D\vec{b},\qquad F\vec{a}=G\vec{b}.\label{eq:matrixform}\ee
where $\vec{a}$ and $\vec{b}$ denote the expansion coefficients $a_{2n,l}$ and $b_{2j,l_1}$ whose indexes are `flattened' in vector notation, $C$, $D$ are matrices whose elements [using \eq{match1}] are 
\be
\begin{array}{c} (C)_{(2n,l),(2j,l_1)}=c_{2n,l}\delta_{n,j}\delta_{l,l_1}\\ \\ (D)_{(2n,l),(2j,l_1)}=d_{(2(n-j),l),(2j,l_1)},\end{array}\ee and similarly for $F$ and $G$ using \eq{match2}. By writing the two equations in block form
\be K\left(\omega \right)\left(\begin{array}{c} {\vec{a}} \\ {\vec{b}} \end{array}\right)\equiv \left(\begin{array}{cc} {C} & {-D} \\ {F} & {-G} \end{array}\right)\left(\begin{array}{c} {\vec{a}} \\ {\vec{b}} \end{array}\right)=0,\ee
the compatibility of the two matching conditions implies the vanishing of (at least) one eigenvalue (or, more generally, singular value in the singular value decomposition) of $K\left(\omega \right)$. A (complex in general) value of $\omega$ compatible with the imposed boundary condition has to be searched, and the corresponding kernel vector then gives the expansion coefficients. In practice it is possible to work with the smaller matrix ($C$ and $F$ would in general be invertible)
\be FC^{-1} D\vec{b}=G\vec{b}\qquad \Rightarrow \left(G-FC^{-1} D\right)\vec{b}=0,\ee
whose kernel vectors give the exterior region coefficients $\vec{b}$, from which $\vec{a}$ immediately follows. The normalization of the wavefunction is discussed in \app{Sec:Normalization}, and in \app{Sec:Derivations3} we lay down for completeness the expansion of integrals which are required in order to calculate expectation values of some general operators (we restrict the expressions to axially symmetric wavefunction with $m=0$).

The constants $S_{2j,l_1}^{\left(a\right)}$ determine the boundary conditions in the asymptotic region. For the partial waves with ${\rm Re}\omega+2j<0$, setting $S_{2j,l_1}^{\left(2\right)}=0$ and $ S_{2j,l_1}^{\left(1\right)}=1$ gives waves exponentially decaying in space (${\rm Im}k>0$), that carry no flux asymptotically. These are the bound, square-integrable components of the wavefunction, that represents the probability density localized to the well. The boundary conditions for ${\rm Re}\omega+2j>0$ depend on whether the problem is to be Hermitian or non-Hermitian.

In order to impose Hermitian boundary conditions with a real quasienergy $\omega$, the terms with $\omega+2j>0$ must include both outgoing and incoming waves. Then we can let $S_{2j,l_1}^{\left(1\right)}=1$ and set $S_{2j,l_1}^{\left(2\right)}$ to the relative phase of waves reflected inwards from the boundary at infinity (assuming that it depends only on the energy and the angular momentum quantum number $l_1$ in the asymptotic, drive-free region). This phase can be used for expansion of the wavefunctions of any Hamiltonian in the region that is far from the scattering center (such as an external particle trap). The solution describes a steady state with a superposition of bound components and traveling waves, incoming and outgoing.

When solving for a complex $\omega$, the boundary condition make the problem non Hermitian. Resonances with Im$\omega>0$ describe a capture process by the oscillating well, with the probability exponentially increasing in time. For the components with ${\rm Re}\omega+2j>0$ both the imaginary and real parts of $k_{2j}$ can be chosen to be positive. Setting $ S_{2j,l_1}^{\left(2\right)}=1$ gives incoming waves whose amplitude exponentially diverges at infinity; the incoming flux of these waves is being captured by the state within the well. Setting $ S_{2j,l_1}^{\left(1\right)}=1$ gives outgoing waves whose amplitude exponentially decays at infinity. 
 Alternatively, resonances with Im$\omega<0$, describing an escape out of the well (exponential decay with time), would have the real part of $k_{2j}$ (for the components with ${\rm Re}\omega+2j>0$) necessarily negative (for the root with positive imaginary part chosen to have the asymptotic decay or divergence as above), which inverts the roles of incoming and outgoing waves; $ S_{2j,l_1}^{\left(2\right)}$ are outgoing waves  diverging at infinity and $ S_{2j,l_1}^{\left(1\right)}$ are incoming waves decaying at infinity. 
 
 The probability density of measuring in the asymptotic region an emitted particle with (real) momentum $\vec{k}$ is given by the squared absolute value of the probability amplitude of the corresponding free particle wavefunction.  Taking into account the interference of the angular harmonics, with different values of $l_1$, at each value of $k_{2j}$, we can define the joint probability density in spherical momentum coordinates
\be f(k,\theta,\varphi)=\frac{1}{\mathcal{N}}\sum_{j}{'}\left|\sum_{l_1}b_{2j,l_1}Y_{l_1}^m(\theta,\varphi)\right|^2\frac{\delta(k-{\rm Re}k_{2j})}{k^2}\label{fktp}\ee
with the summation index $j$ extending over all outgoing channels that do not decay asymptotically. The normalization of \eq{asymptoticj} guarantees that each channel is weighted correctly, and $\mathcal{N}$ sets the overall normalization.
The probability to measure a particle with momentum in the volume element of momentum space between $(k,\theta,\varphi)$ and $(k+dk,\theta+d\theta,\varphi+d\varphi)$ is then
\be p(k,\theta,\varphi)=f(k,\theta,\varphi)k^2\sin\theta dkd\theta d\varphi,\label{eq:pkthetaphi}\ee
 with the normalization $\int p = 1$ determining the value of $\mathcal{N}$. We can also define the axially symmetric marginal probability density
\be f(k,\theta)=\int d\varphi f(k,\theta,\varphi),\ee
that will be used in the following. From $f(k,\theta)$ it is a simple change of coordinates to $f(k_\rho,k_z)$ which is the joint distribution in terms of axial and transverse momentum, from which the marginal distributions can also be obtained.
 
 %For $\mathfrak{Im}\left(\omega\right)>0$, and k with a positive imaginary part, the $H^\left(1\right) are outgoing waves with a decaying amplitude at infinity. The time evolution is a build up.
%If we take $\mathfrak{Im}\left(\omega\right)<0$, and choose k to be the root with positive imaginary part, its real part is necessarily negative, and inverses the sense of incoming/outgoing.  $H^\left(1\right) are incoming waves whose amplitude is 0 at infinity, $H^\left(2\right)$ are outgoing waves whose amplitude explodes at infinity. The time evolution is decaying.

\subsection{Scattering wavefunctions}\label{Sec:Scattering}

In a scattering problem the imposed boundary conditions are composed of a given form of free particles waves in the `input' channel, with a well-defined energy value $\omega>0$, in the asymptotic region. The given wave is interacting with the potential within its range of affect, resulting in superimposed scattered outgoing spherical waves in the asymptotic region. We will pose the boundary conditions in terms of ingoing spherical waves, that facilitates the calculation of the unitary S-matrix. 

Specifying the values of $S_{2j,l_1}^{\left(2\right)}$ in \eq{phi3DS2S1} and treating $S_{2j,l_1}^{\left(1\right)}$ as free parameters, the matching condition becomes the inhomogeneous linear equation
\be \left[G^{(1)}-FC^{-1}D^{(1)} \right]\vec{S}^{\left(1\right)}= -\left[G^{(2)}-FC^{-1}D^{(2)} \right]\vec{S}^{\left(2\right)}.\ee 
Setting $S_{2j,l_1}^{\left(2\right)}= \delta_{j,j'}\delta_{l_1,l_1'}$ describes an incoming spherical wave with asymptotic unit flux in the channel $(j',l_1')$, as the normalization of \eq{scatteringnormalization} makes $\left|\vec{j}\right|=\left|S_{2j,l_1}^{\left(a\right)}\right|^2$ in each channel. The resulting $S_{2j,l_1}^{\left(1\right)}$ gives the  $\left[(j,l_1),(j',l_1')\right]$ matrix-element
of the unitary scattering S-matrix that transforms an incoming wave to an outgoing.
To simplify the expressions in the following, we further assume that the incoming wave is purely s-wave in a single energy channel, i.e. $(j',l_1')=(0,0)$. This corresponds to the limit of scattering of slow particles, if the energy is also low enough \cite{landau1981quantum}. Then the elastic  cross-section is
\be 
\sigma_{e,0} / (2\pi) = \left|1 - S_{0,0}\right|^2 / (4\omega),\label{sigmae0}
\ee
 the inelastic cross-section is given by
\be
\sigma_{r,0} / (2\pi) = (1 - \left|S_{0,0}\right|^2) / (4\omega),\label{sigmar0}
\ee
and the total cross-section is \be \sigma_{t,0} / (2\pi) = 2(1 - {\rm Re}\,S_{0,0}) / (4\omega),\label{sigmat0} \ee
where we have removed the superscript from $S_{0,0}^{(1)}$ to simplify the notation.
We note that for scattering solutions, time-reversal (with complex conjugation) interchanges the initial and final states and reverses the direction of wave propagation, and for a Hamiltonian which is invariant, the scattering amplitude must remain the same -- this is the reciprocity theorem.

\subsection{Analytic properties of the solutions}\label{Sec:Analytic}

The asymptotic form of $V_{\rm int}$ determines important properties of the scattering solutions, bound states and poles of the S-matrix in complex energy and momentum planes. In this subsection we review a few of these properties \cite{landau1981quantum,burke2011r,newton2013scattering}, that will be used in the following sections.
We will restrict the discussion to two forms for the interaction potential; either $V_{\rm int}$ that vanishes identically beyond a certain distance (a finite range potential), or is asymptotically an (attractive) power law potential $V_{\rm int} \sim -C/r^\alpha$ with $C>0$ and $\alpha>3$ which is a restricted form of what is typically referred to as a short range (or `shorter-ranged') potential \cite{friedrich2017theoretical}. 

If we consider the time-independent Schr\"{o}dinger equation (with potential $V_{\rm int}$), we can write its solutions in the form
\be \phi_{\varepsilon lm} \left(\vec{r}\right)=\frac{1}{r} u_{\varepsilon l} \left(r \right)Y_{l}^{m} \left(\theta ,\varphi \right),\label{phi1overruY}\ee
where $u_{\varepsilon l} \left(r \right)$ are solutions of the reduced equation
\be \left[-\frac{1}{2}\frac{d^{2} }{dr ^{2} } +\frac{l\left(l+1\right)}{2r ^{2} } +V_{\rm int}\left(r\right) -\varepsilon \right]u_{\varepsilon l} \left(r \right)=0\label{eqradial}.\ee
At each (complex) value of $\varepsilon$, \eq{eqradial} has two linearly independent solutions, $u_{\varepsilon l}^{(1,2)}$. If the potential at $r\to\infty$ behaves like $-C/r^{\alpha}$ with $\alpha>2$ then the two linearly independent solutions $u_{\varepsilon l}^{(1,2)}$  can be chosen to have an $\varepsilon$-independent limit at $r\to 0$ for all energies. In addition, the potential cannot have an infinity of bound states accumulating near the threshold $\varepsilon=0$. For $\alpha > 3$, $u_{\varepsilon l}^{(1,2)}$  are entire functions \cite{Gao2008}, of the complex momentum $k$ or of the energy (on its two-sheeted Riemann surface, with the cut extending on the positive real axis), at any fixed $r$. This property, that 
holds also for the solutions of a finite range potential, will be alluded to in \seq{Sec:Outlook}. When this property fails, $u_{\varepsilon l}^{(1,2)}$ will not be analytic in the entire $k$ plane, only in parts of it, and there may exist cuts on the imaginary $k$ axis (at negative energy).

At the same time, a potential with a tail with $\alpha>2$ cannot keep the same power-law behavior down to the origin, because then the energy spectrum will not be bounded from below (due to an infinity of bound states at decreasing energies). Hence physically relevant short range atomic potentials will rise near the origin, supporting a finite number of bound states, which is again true for finite range potentials as well.
The physical solution to \eq{eqradial} is defined to be the unique linear combination that is regular at the origin, normalized by a definite condition, e.g. $r^{-l-1}u_{\varepsilon l}\to 1$. Since this boundary condition is $\varepsilon$- (and \,$k$-) independent, the physical solution is an entire function of $\varepsilon$ ($k$) for any fixed $r$.
When both of the functions $u_{\varepsilon l}^{(1,2)}$ are entire, this immediately implies that the scattering matrix is meromorphic in complex energy (or momentum space) with isolated poles. Poles on the imaginary $k$ axis correspond to bound states, and there are no other poles in the upper momentum half-plane (i.e. poles in the physical energy sheet can only be bound states with negative energy). If the S-matrix has a pole at $k$, then it also has a pole in $-k^*$, and zeros at $-k$ and $k^*$. Hence a bound state actually hides two coinciding poles. In \fig{Fig:OmegaSpace0} and in \fig{Fig:Exceptional} of \seq{Sec:Outlook} it can be clearly seen how these two poles separate at the presence of the periodic perturbation.

Finally, the partial wave series of the scattering solution is convergent and the scattering amplitude (and cross sections) finite at all angles for a potential $V_{\rm int}$ that decreases asymptotically faster than $1/r^{3}$. In that case, the scattering in the limit of low velocity ($\omega\to 0$) is isotropic and independent of the energy. This can be expressed using the s-wave scattering length $a$ (that is well defined), and the fact that $\sigma_e\to 4\pi \left|a\right|^2$. At the presence of inelastic interactions, the scattering length is not necessarily real but can have a (negative) imaginary part that gives the inelastic cross section, $\sigma_r\to 4\pi\left|{\rm Im}a\right|/k$, that is inversely proportional to the velocity in the low velocity limit (the $1/v$ law). These limiting behaviors of the cross sections will be shown to hold in \fig{Fig:s-waveimages2} and \fig{Fig:p-wavecurves}.

\begin{figure}[ht]
\center {\includegraphics[width=3.4in]{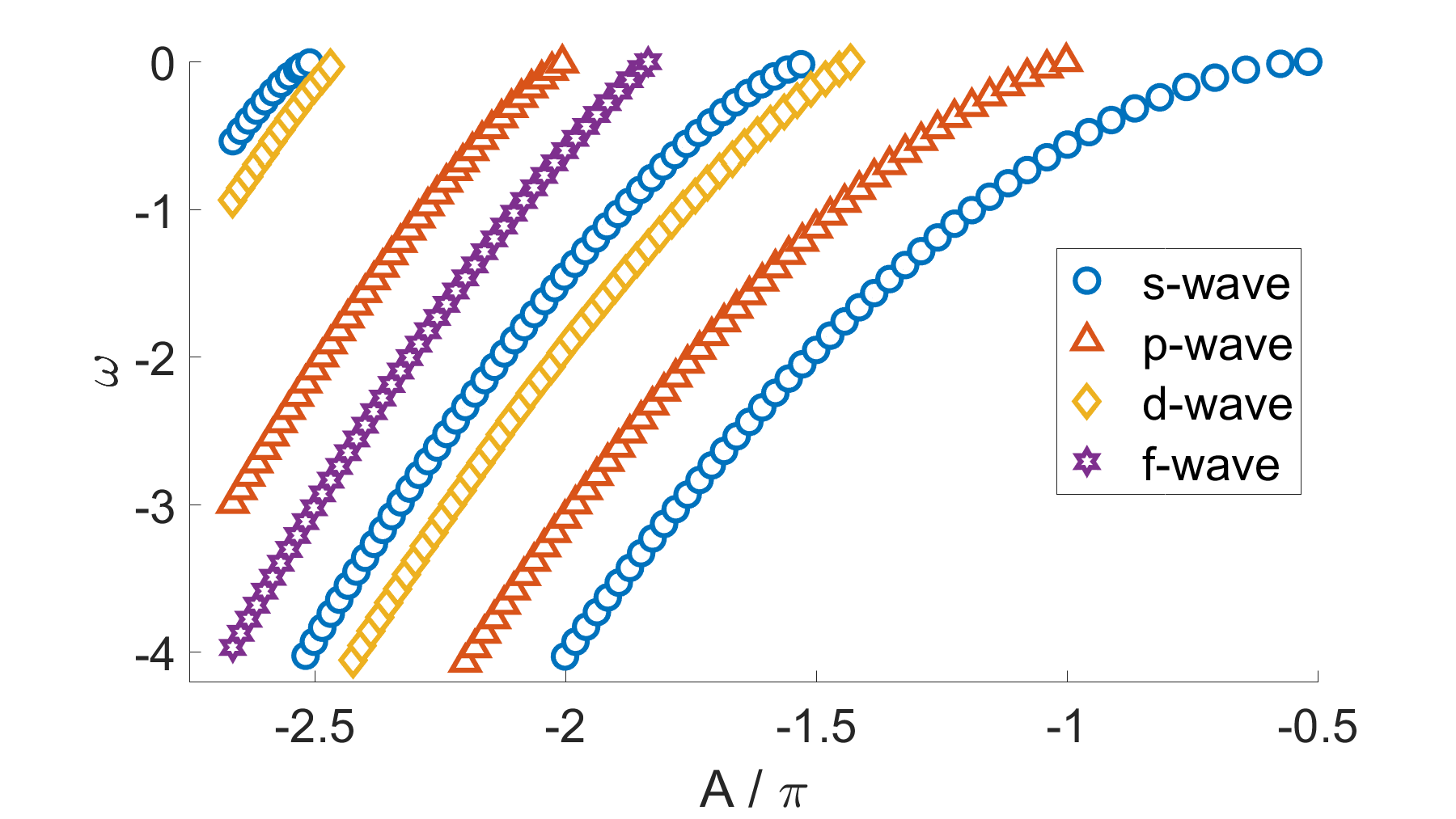}
\caption
{The spectrum of the square well potential [\eq{squarewell}] as a function of the universal parameter $A$ [\eq{eq:A}]. As  $-A/\pi$ crosses an integer (half-integer) value, a new bound p-wave (s-wave) appears in the well. Higher order $l$ states appear progressively within these intervals. The nondimensional energy scale is fixed  by setting the well size to $d=2$.
\label{Fig:Spectrum}}}
\end{figure}

\section{Poles and Singularities with Periodic Driving}\label{Sec:Numerics}

In this section we employ the methods presented in the previous sections to study a model system consisting of a spherically-symmetric square-well potential and a time-dependent periodic linear drive which acts outside of the well.
Using the fundamental frequency of the periodic drive, $\Omega$, we can define the length and energy scales
\be d_{o} =\sqrt{2\hbar /m \Omega }, \qquad E_o = \hbar\Omega/2, \label{units}\ee
and originally dimensional variables become nondimensional by rescaling according to 
\be \vec{r}\to \vec{r}/d_{o}, \qquad \vec{k}\to \vec{k}d_{o}, \qquad t\to t\Omega /2,\ee
after which we have explicitly $\hbar=m=1$ and the drive's frequency in these units is $\Omega=2$. With a spherical square-well potential,
\be V_{{\rm int}} \left(\left|\vec{r} \right|\right)=\left\{\begin{array}{cc} {-V_{0} ,} & {\left|\vec{r} \right|<d} \\ {0} & {\left|\vec{r} \right|>d} \end{array}\right.\label{squarewell}\ee
(where $d$ and $V_0$ are nondimensional, measured in the units of \eq{units}), 
and the regular solution inside the well is a spherical Bessel function,
\be \phi _{\left\{k,l,m\right\}} \left(\vec{r},t\right)\propto e^{-i\left(\frac{1}{2} k^{2} -{V_{0} } \right)t}j_{l} \left(kr\right)Y_{l}^{m}.\ee
In those units, we take the periodic force of \eq{Fpi} to be a simple harmonic drive with amplitude $F_2$,
\be F^{\pi}\left(t\right) = F_2\cos2t,\label{eq:F2}\ee
regaining \eq{eq:F0}.

We define the well parameter 
\be A=-\sqrt{2V_0d^2},\label{eq:A}\ee that is the single parameter that characterizes the square well \cite{nussenzveig1959poles}. 
In Fig.~\ref{Fig:Spectrum} we show the spectrum of bound states of the time-independent square-well over a small range as function of the universal parameter $A/\pi$. The energy scale is obtained using \eq{units}, fixing $d=2$. It can be immediately seen from \eq{eq:A} that varying $\Omega$ is equivalent to leaving $A$ invariant while scaling both $V_0$ and $d$, as the  energy and distance units are rescaled.

\begin{figure}[t!]
\center {\includegraphics[width=3.1in]{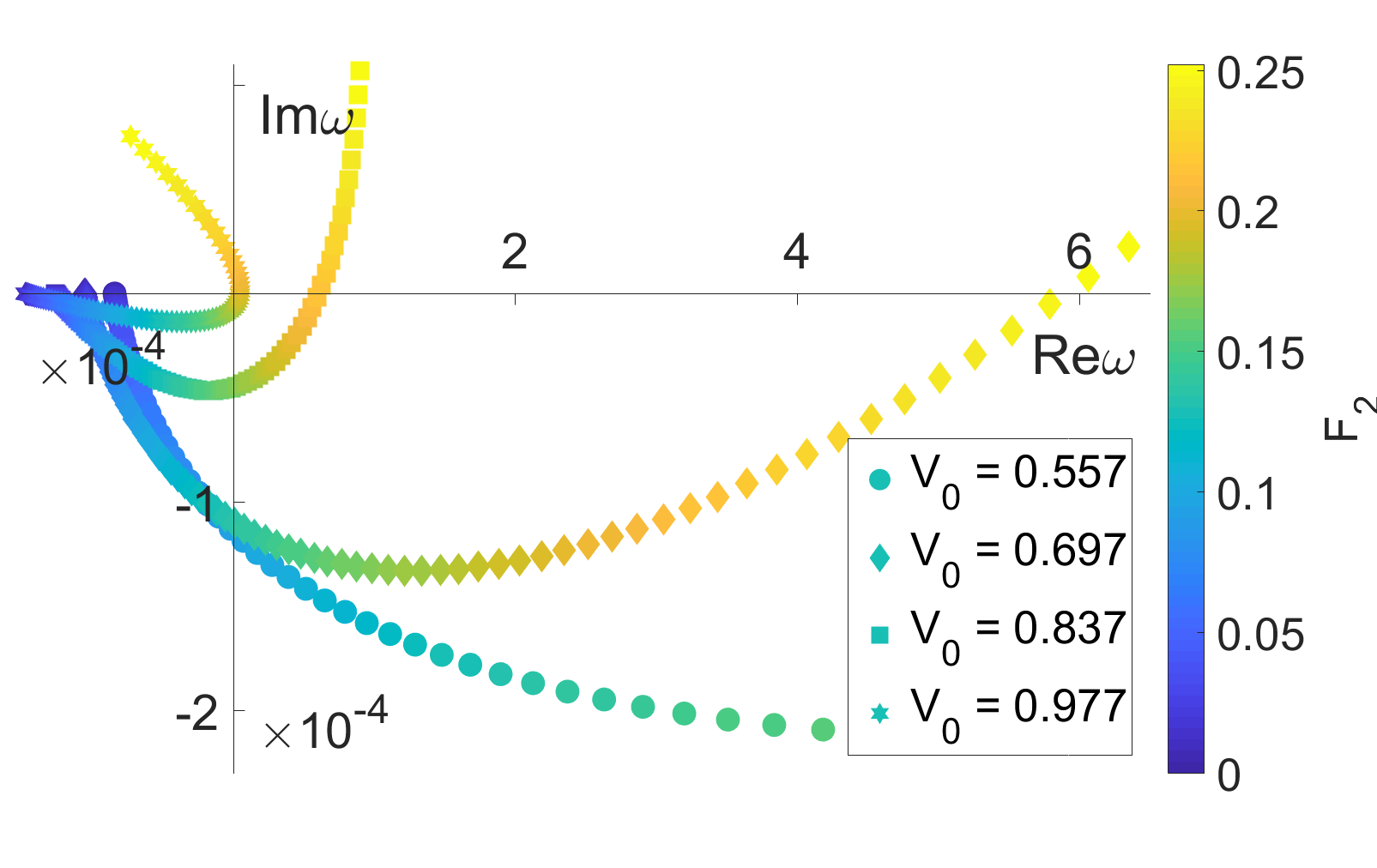}
\caption
{Complex $\omega$-space showing 4 resonances of the square-well potential [\eq{squarewell}] followed as a function of $F_2$ [\eq{eq:F2}] by solving \eqs{iphidotLR2}-\eqref{Vin1Vout1}, i.e., the drive is restricted to act outside the well (in contrast to \fig{Fig:OmegaSpace0}). The well parameter of \eq{eq:A} is fixed at $-A/\pi\approx 0.504$, with $V_0$ taking 4 different values shown in the legend, $d$ being adjusted accordingly. Starting from the (single) loosely-bound s-wave state (see \fig{Fig:Spectrum}), the  poles of the S-matrix correspond initially (while Im$\omega<0$) to the escape process (emission out of the well). The pole for $V_0=0.557$ crosses the real $\omega$-axis outside of the figure boundaries, at the critical parameters [\eq{Fomegac}] $\bar{F}^c\approx 0.260$, $\bar{\omega}^c\approx 3.35 \times 10^{-3}$ (see \fig{Fig:Rate1} to \fig{Fig:s-waveimages2}).
\label{Fig:OmegaSpace1}}}
\end{figure}

\subsection{Driving a loosely bound s-wave}\label{Sec:Looselys-wave}

In this subsection we set $-A/\pi\approx 0.504$, i.e.~the well is shallow with a single bound s-wave state close to threshold. We solve the problem by plugging the potential [\eq{squarewell}] into \eq{Vin1Vout1}, i.e., the drive is restricted to act outside the well. This is in contrast to the results presented in \fig{Fig:OmegaSpace0}, where the drive was solved for in all space. Choosing this form is motivated by the fact that for a general interaction potential, the problem cannot be solved exactly together with the periodic drive, and some sort of approximation is required. A possible choice [introduced in \eq{iphidotLR}] is to divide space into two regions where either the interaction or the periodic drive act. A further discussion and comparison of the two models will be presented in \seq{Sec:Approximation}.

\begin{figure}[t!]
\center {\includegraphics[width=3.1in]{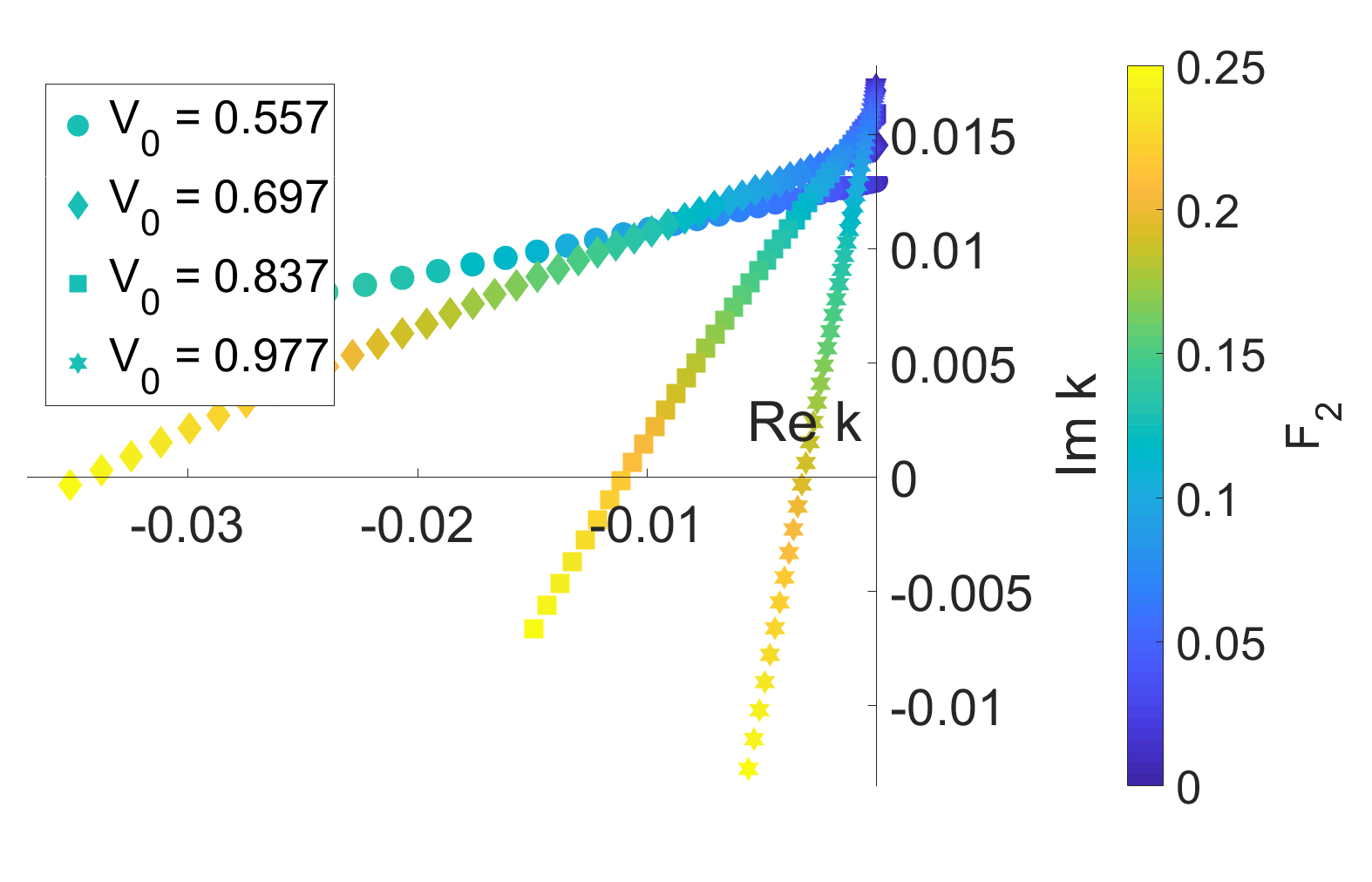}
\caption
{Complex $k$-space showing the value of $k_0$ at the poles of \fig{Fig:OmegaSpace1}. The pole trajectories can be seen to lie almost along straight lines in this case (see however \fig{Fig:kSpace2}). The poles of the corresponding capture process (related by time-reversal invariance of the Hamiltonian) have mirror-imaged trajectories at the right half-plane (not shown). For each (capture or emission) pole of the S-matrix, there is also a zero at $-k$.
\label{Fig:kSpace1}}}
\end{figure}

\begin{figure}[ht] \center {\includegraphics[width=3.0in]{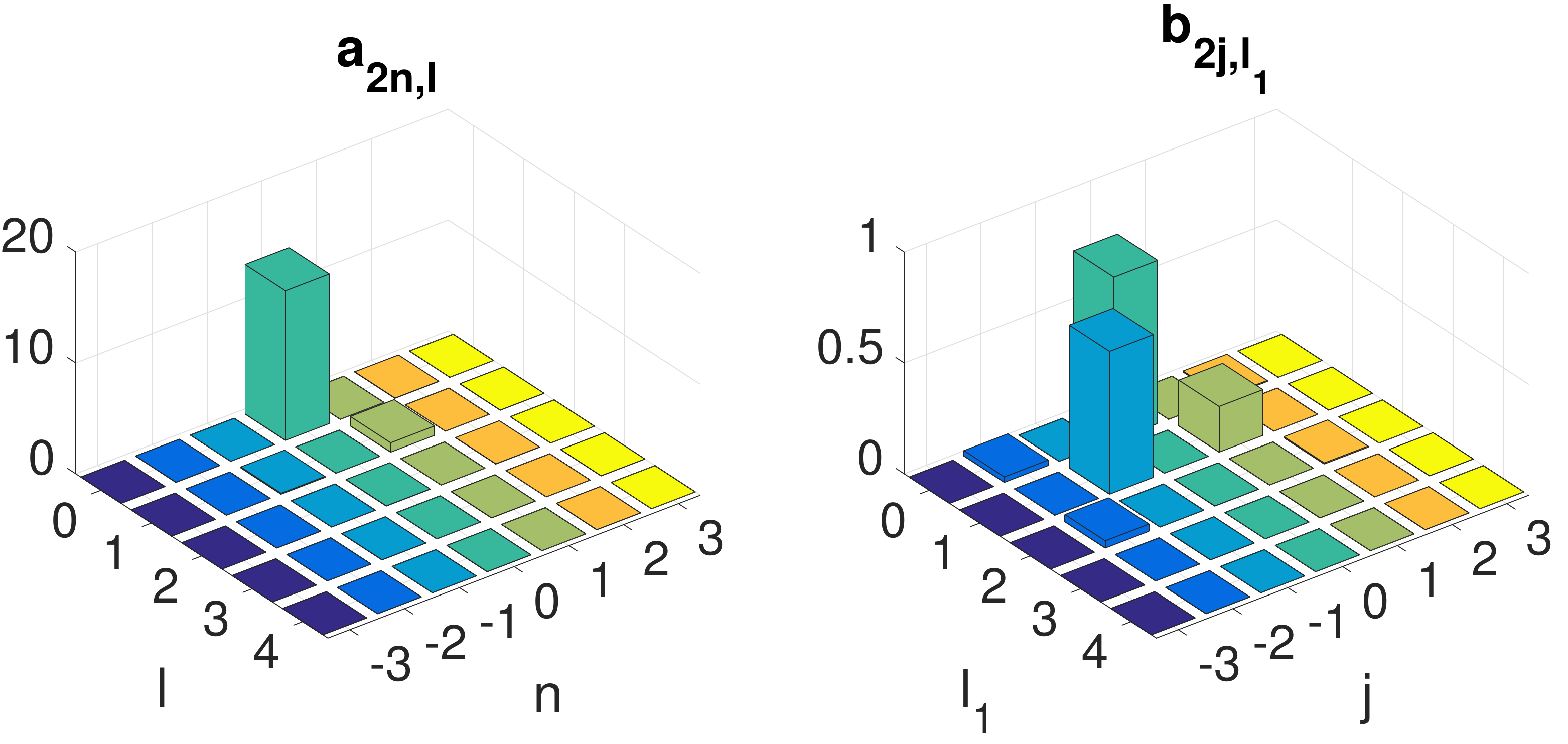}\caption {Absolute value of the matching coefficients of the solution [\eq{phi3Dansatz}], in the interior region ($a_{2n,l}$) and in the exterior region ($b_{2j,l_1}$), for the lowest pole  \fig{Fig:OmegaSpace1} with $V_0=0.557$, at ${F}_2= 0.03$. The superposition of (a small number of) components can be seen, both inside and outside the well, of decaying ($j\le 0$) as well as travelling waves ($j\ge 1$).\label{Fig:MatchingCoefficients}}}\end{figure}

Figure \ref{Fig:OmegaSpace1} shows the values of $\omega$ for the (initially) escape (radiating) pole followed by continuation from the s-wave bound state at $F_2=0$, at 4 different values of $V_0$ (with $A$ fixed). It can be seen that the crossing of the real $\omega$ axis is generic and can be realized at different values of the parameters. 
For low drive amplitude, the quasi-bound state's decay rate grows quadratically (as can be inferred from a log-log plot, not shown), which is the expected perturbative result [\eq{epsilon2n}]. In the nonperturbative regime the  decay rate is clearly nonmonotonous; for a strong enough drive it decreases and reaches 0, as the poles reach the real $\omega$-axis.

\begin{figure}[ht]
\center {\includegraphics[width=3.2in]{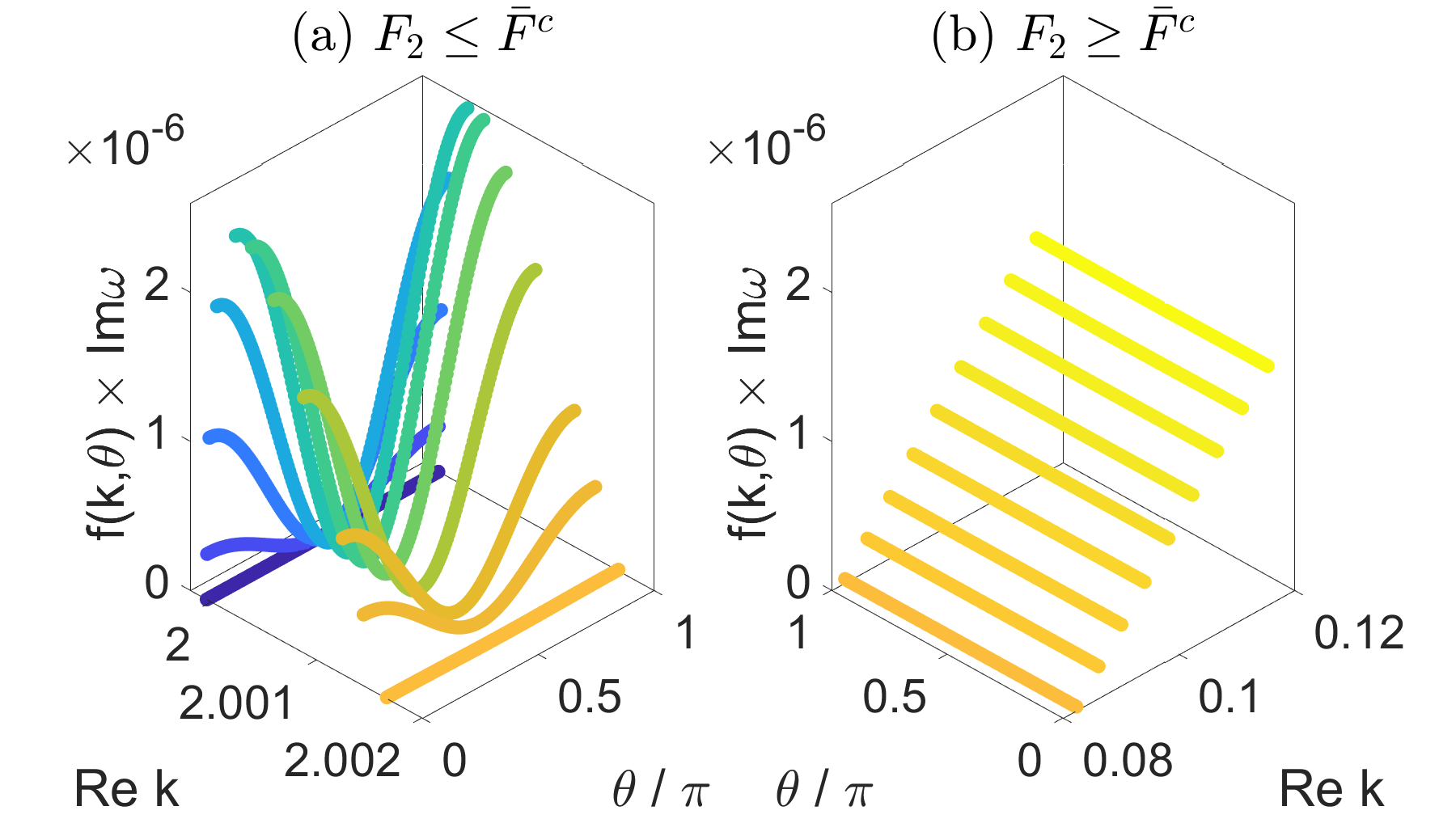}
\caption
{The leading order contribution to the the probability density $f(k,\theta)$ in spherical coordinates of momentum flux [see \eq{eq:pkthetaphi}], multiplied by (half) the rate of emission out of the well, plotted for increasing values of $F_2$, for the lowest pole in \fig{Fig:OmegaSpace1} with $V_0=0.557$. The curve colors (from dark blue to bright yellow) encode the value of $F_2$. (a) For $F_2$ up to the critical value $\bar{F}^c\approx 0.260$ the radiation is mostly p-wave in the $j=1$ channel. (b) At the critical value the  distribution switches abruptly to the $j=0$ channel (mostly s-wave). The multiplication by the emission rate that goes through 0 makes the physical process smoothly decaying and then rising, despite the fact that the distribution $f(k,\theta)$ changes abruptly.
\label{Fig:Rate1}}}
\end{figure}

Figure \ref{Fig:kSpace1} shows the momentum space values of $k_0$ for the same solutions, lying on nearly straight lines . The slope of the lines varies continuously with the parameters. The properties of the solutions were described at length in \seq{Sec:Intro}, and  we further discuss the parametric dependence in \seq{Sec:Outlook}.
Since in momentum space, the poles that start on the imaginary $k$-axis must cross the bisector of their quarter plane before reaching the Im$k=0$ line, in $\omega$ plane the poles follow a curved trajectory around the origin, crossing to Re$\omega>0$. This guarantees that after the critical point, the solutions that have switched roles (between capture and emission) remain valid -- with Im$\omega\neq 0$ and one channel ($j=0$) that doesn't decay asymptotically, as required by the continuity equation.

The superposition of different bound and diverging components can be seen in the  solution coefficients of the expansion in \eq{phi3Dansatz}, which are depicted in \fig{Fig:MatchingCoefficients} for a particular state. The quasi-bound s-wave state which for $F_2=0$ would have its entire amplitude at $\left(n=0, l=0\right)$ and $\left(j=0, l_1=0\right)$, has developed a superposition of partial waves (here mostly outside of the well). The `checkerboard' pattern is the result of the dipolar nature of the coupling, which conserves $\left(-1\right)^{n+l}$ [or $\left(-1\right)^{j+l_1}$], and can be used in practice to speed up the numerical calculations.
Figure \ref{Fig:Rate1} shows the radiation pattern in the leading nondecaying channel for the lowest radiating pole of \fig{Fig:OmegaSpace1} with $V_0=0.557$. The joint probability density of the radiated waves given by $f(k,\theta)$ of \eq{eq:pkthetaphi} multiplied by Im$\omega$, is plotted at discrete steps of $F_2$ which determines the value of Re$k$. On the left, for $F_2$ lower than the critical value $\bar{F}^c\approx 0.260$, the radiation is an odd Legendre polynomial of $\cos\theta$ (mostly p-wave) in $j=1$ channels, with the flux initially increasing and then decreasing. On the right hand side, immediately after the pole crosses the real $\omega$ axis (at $\bar{\omega}^c\approx 3.35\times 10^{-3}$), the radiation abruptly collapses into the $j=0$ channels, composed mostly of s-waves (and other even harmonics), with the flux increasing as $F_2$ further increases. 

\begin{figure}[ht]
\center {\includegraphics[width=3.3in]{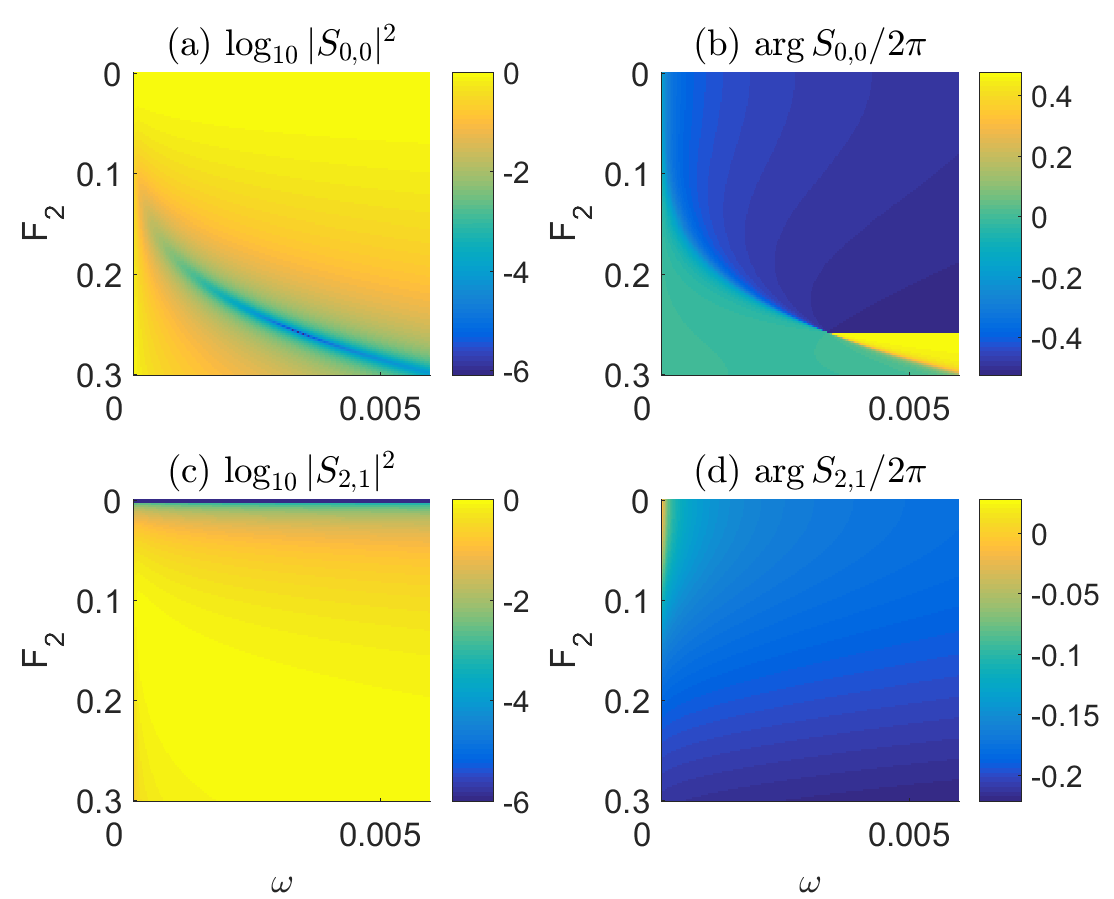}
\caption
{(a) The squared absolute value of the S-matrix element of the incoming s-wave ($|S_{0,0}|^2$), color coded on a logarithmic scale as function of $F_2$ and the real incoming wave energy $\omega$. $S_{0,0}$ vanishes at the value of $F_2$ and energy for which the pole lies on the real energy axis  ($\bar{F}^c\approx 0.260$, $\bar{\omega}^c\approx 3.35 \times 10^{-3}$, for the same pole as \fig{Fig:Rate1}), with the finite grid size of the figure giving a small nonzero value. This corresponds to total absorption in this channel -- all incoming waves are scattered into the other output channels. (b) The argument of the S-matrix element, ${\rm arg}\,S_{0,0}$, showing that a $2\pi$ phase is accumulated around the critical point in this parameter space. (c) The amplitude of outgoing p-waves in $j=1$ channel, given by $|S_{2,1}|^2$ that approaches unity near the critical point (in a logarithmic scale). (d) The argument ${\rm arg}\,S_{2,1}$, which shows no feature like that of panel (b).
\label{Fig:s-waveimages}}}
\end{figure}

\begin{figure}[ht]
\center {\includegraphics[width=3.3in]{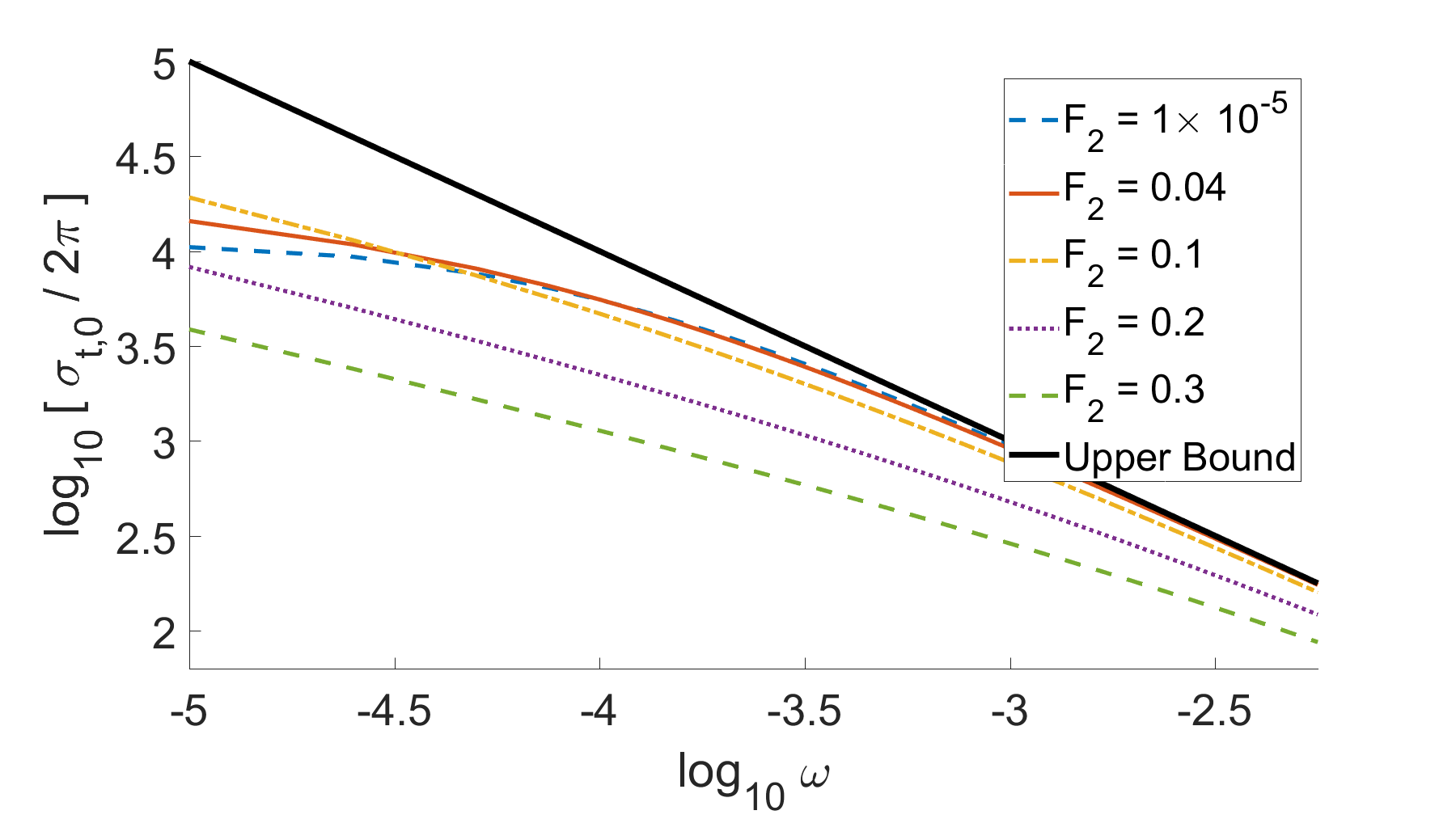}
\caption
{The base-10 logarithm of the  total scattering cross sections divided by $2\pi$ [\eq{sigmat0}], as a function of the real incoming wave energy $\omega$, for the input s-wave channel (which gives the scattering in the limit of slow particles). The parameters are identical to \fig{Fig:s-waveimages}. The curves at several $F_2$ values, plotted on a log-log scale, show how for $F_2\to 0$ the scattering is predominantly elastic (with the $\omega\to 0$ dependence being that of $\sigma_{e,0}$, i.e.~is energy-independent), and becoming increasingly dominated by the inelastic scattering ($\propto \omega^{-1/2}$). The cross section does not show any nonmonotonous features however -- compare with \fig{Fig:p-waveimages} and \fig{Fig:p-wavecurves}.
\label{Fig:s-waveimages2}}}
\end{figure}

Figure \ref{Fig:s-waveimages} shows some quantities of the S-matrix of a scattering formulation as a function of $F_2$ and the energy $\omega\ll 1$ of an incoming s-wave (which determines the limit of scattering of slow particles, see \seq{Sec:Scattering}). The parameters are the same as for the pole followed in \fig{Fig:Rate1}. The critical point [defined in \eq{Fomegac}] where the two complex conjugate poles (and zeros) of the S-matrix coincide on the real energy axis can be identified as at this point the S-matrix element $S_{0,0}$ vanishes (making the argument undefined), and a $2\pi$ phase is accumulated if going around this point in $(F_2,\omega)$ parameter space. ${\rm arg}\,S_{0,0}$ is defined by continuity from $\omega\to 0$ for each fixed value of $F_2$. Along such a line for which $F_2<\bar{F}^c$ there is a sharp decrease of the argument as function of $\omega$, while for  $F_2>\bar{F}^c$ there is a sharp rise. The S-matrix element $S_{2,1}$ of scattered p-waves in the $j=1$ channel approaches unit modulus (in a large region of the parameter plane). At the critical point there is total absorption of the incoming s-wave, and it is being radiated out as a (mostly) p-wave, with energy higher by at least a drive quantum. 

In \fig{Fig:s-waveimages2} the total  scattering cross section is  plotted for s-waves with $\omega\ll 1$. The features of scattering of slow particles discussed in \seq{Sec:Analytic} (the energy-independent elastic scattering length and the $1/v$ law for inelastic scattering) hold, as can be deduced from an examination of the elastic and inelastic cross section curves (not shown). The total cross section changes its behaviour as a function of the drive amplitude -- for $F_2\to 0$ the scattering is  elastic, and it becomes increasingly inelastic as $F_2$ is increased. The cross section however is monotonous as a function of the energy (as is typical for an s-wave resonance) throughout the large variation of $F_2$. This is very different from the results of driving a bound p-wave state to be discussed in the next subsection.

\subsection{Driving a deeper bound p-wave}\label{Sec:p-wave}

\begin{figure}[t!]
\center {\includegraphics[width=3.0in]{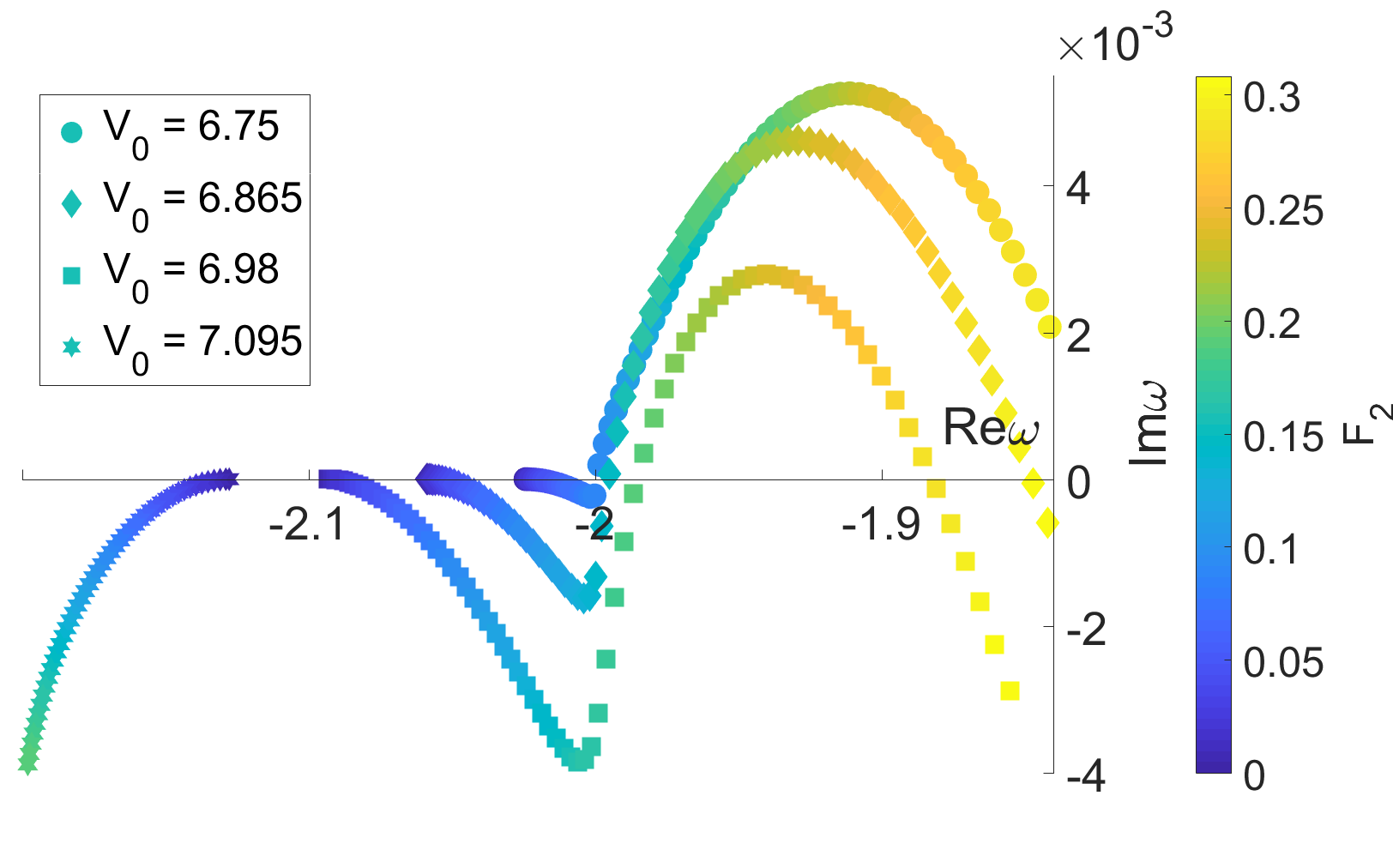}
\caption
{Complex $\omega$-space (as in \fig{Fig:OmegaSpace1}) showing 4 resonances of the square-well potential for the well parameter fixed at $-A/\pi\approx 2.565$. Each being initially a p-wave bound state with energy $\omega(F_2=0)\lesssim -2$, at least $N\ge 2$ quanta of the external drive (of frequency $2$) must be absorbed in each radiating channel (Re$\omega +2j>0$, $j\ge 2$). Even as the three S-matrix poles which are pushed towards a larger quasi-energy cross to Re$\omega\ge -2$, the same channels ($j\ge 2$) are the only radiating ones, since the $j=1$ channel remains incoming and decaying. At the crossing of the real $\omega$-axis the solutions are singular in a similar manner to the near-threshold s-waves.
\label{Fig:OmegaSpace2}}}
\end{figure}

\begin{figure}[ht]
\center {\includegraphics[width=3.3in]{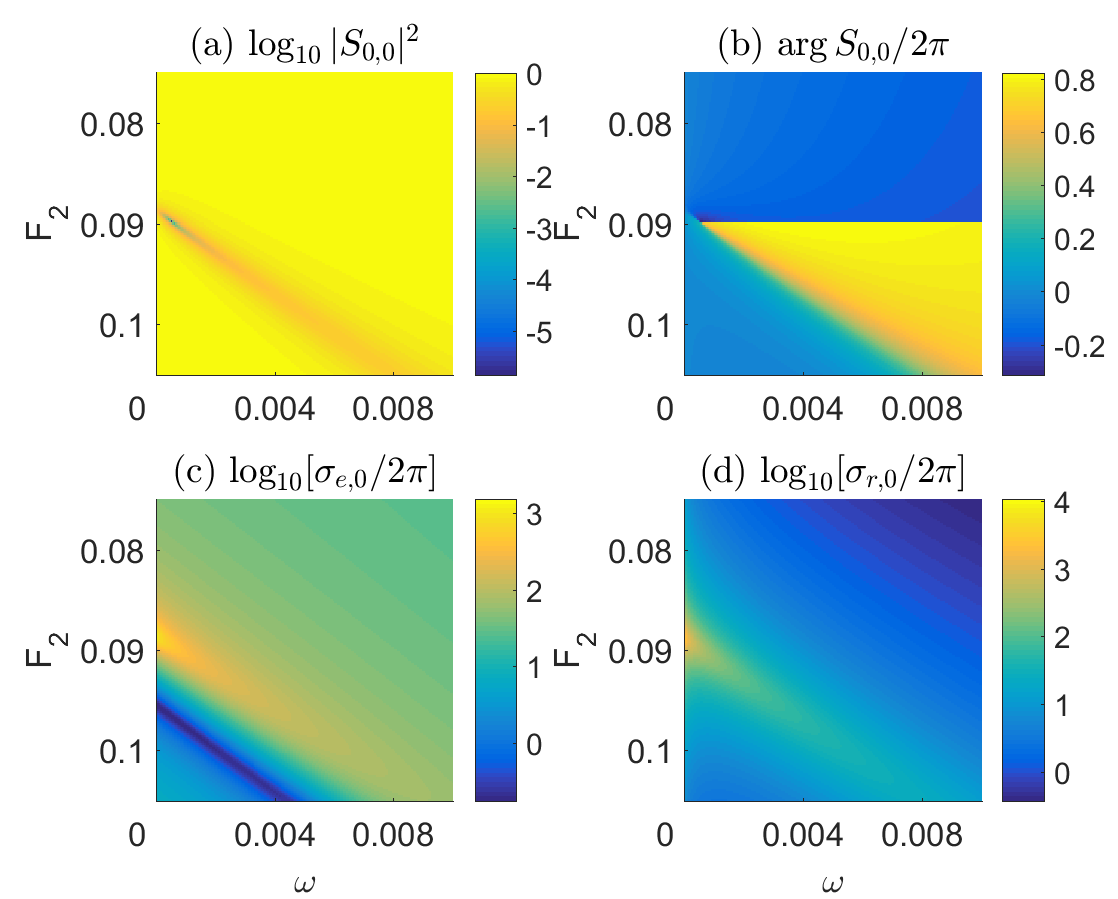}
\caption
{(a) As in \fig{Fig:s-waveimages}, $|S_{0,0}|^2$ color coded on a logarithmic scale as function of $F_2$ and $\omega$, corresponding to the pole with $V_0=6.75$ in \fig{Fig:OmegaSpace2} (with $\bar{F}^c\approx 0.0898$, $\bar{\omega}^c\approx -1.9995$). (b) The argument of the S-matrix element, ${\rm arg}\,S_{0,0}$, showing a $2\pi$ phase jump, around $\bar{F}^c$ and $\omega\approx 5\times 10^{-4}$, which plausibly corresponds to an incoming s-wave completely transferred into the quasi-bound p-wave state by emitting a quantum of energy into the drive, and then being radiated as a p-wave, 2 quanta higher in energy. We note the different range of the argument as compared with \fig{Fig:s-waveimages}, determined by the $\omega\to 0$ limit. (c) The base-10 logarithm of the elastic [\eq{sigmae0}] and (d) the inelastic [\eq{sigmar0}] cross sections divided by $2\pi$, for the input s-wave channel (in the limit of scattering of slow particles). Sharp features exist in the cross sections (more details are discernible in \fig{Fig:p-wavecurves}) in a relatively small range of $F_2$ and a large range of energy, to be contrasted with \fig{Fig:s-waveimages2}. 
\label{Fig:p-waveimages}}}
\end{figure}

\begin{figure}[ht]
\center {\includegraphics[width=3.2in]{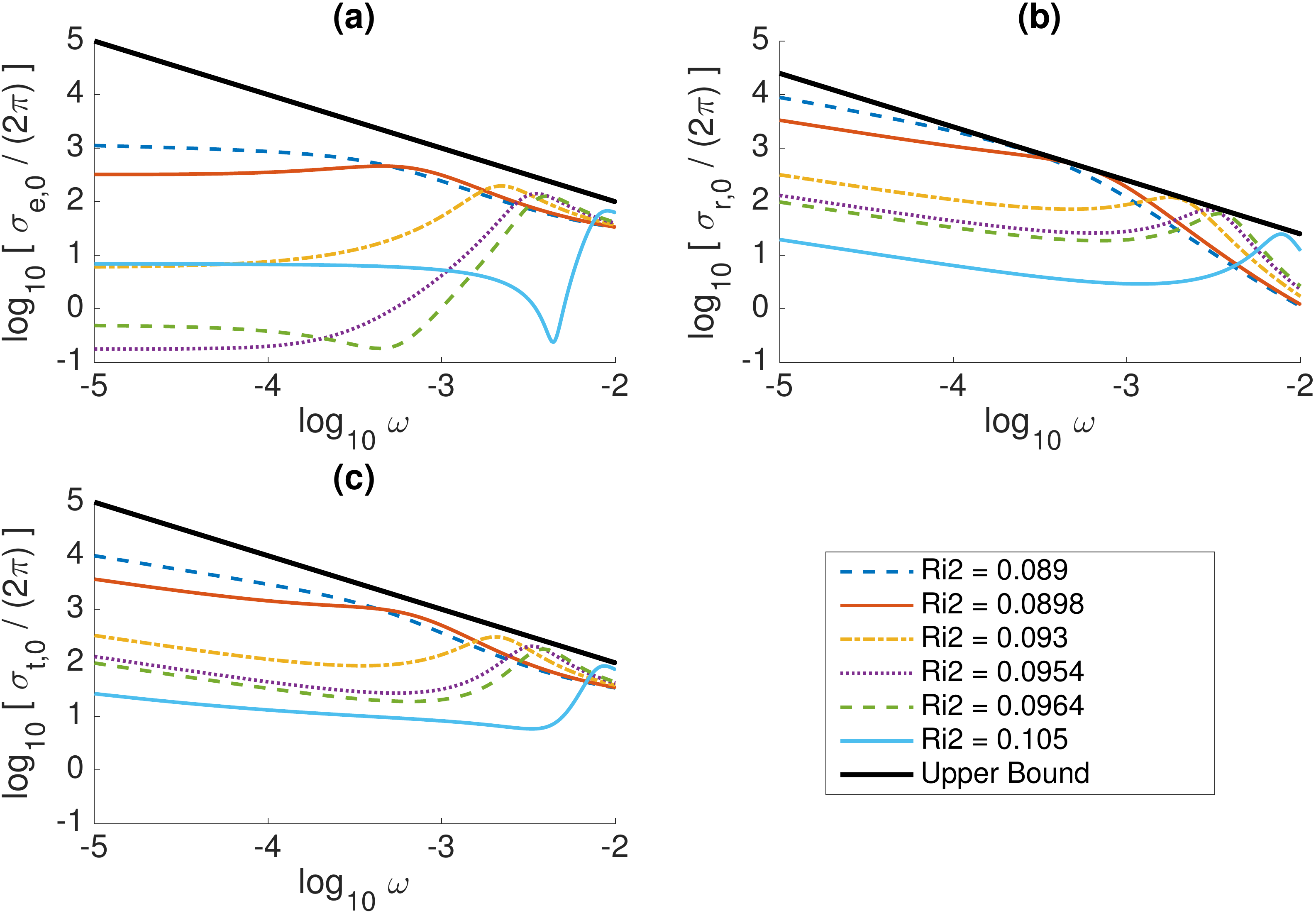}
\caption
{(a) The elastic, (b) inelastic, and (c) total scattering cross sections divided by $2\pi$ [\eqs{sigmae0}-\eqref{sigmat0}], vs.~the energy of the incoming wave (on a log-log plot), taken at a few values of $F_2$ around the critical value $\bar{F}^c\approx 0.0898$, for the same parameters as in \fig{Fig:p-waveimages}. The nonmonotonous features of the cross sections as function of energy, and the strong alteration of the curves as function of $F_2$ are evident, as well as the $\omega\to 0$ dependence of $\sigma_{e,0}$ (energy-independent) and of  $\sigma_{r,0}$ ($\propto \omega^{-1/2}$), as required by the slow-particle scattering limits.
\label{Fig:p-wavecurves}}}
\end{figure}

In this subsection we set $-A/\pi\approx 2.565$ and focus on a p-wave (with magnetic quantum number $m=0$) whose binding energy (for $F_2=0$) is a little larger than $2$ -- absorption of two quanta is necessary (for $F_2 \ll 1$) to emit outgoing waves. The same equations are solved as in the previous subsection.
Figure \ref{Fig:OmegaSpace2} shows the values of $\omega$ for four poles followed by continuation. Three of those poles cross towards Re$\omega>-2$ while there is one pole that is pushed towards negative energies. An exceptional point, discussed further in \seq{Sec:Outlook}, separates the poles going left and right.
As in the previous subsection, as the poles in the lower half plane cross the line Re$\omega= -2$, the channels whose energy becomes positive (and for that could be termed `open') remain in fact asymptotically decaying (and in fact incoming) so the radiation pattern does not show a qualitative change. 

As the poles cross to the upper half $\omega$-plane, the solutions change their nature abruptly and they are singular on the real energy line. The pole trajectories in momentum space will be shown in \seq{Sec:Outlook} and present a nontrivial behavior. The radiation pattern (not shown) presents similar features to that of \fig{Fig:Rate1} (with the required modifications of the momentum values and the distribution shape).

Figures \ref{Fig:p-waveimages}-\ref{Fig:p-wavecurves} show the characteristics of a scattering setup with slow particles. The limiting low energy behaviors of the cross sections, discussed in \seq{Sec:Analytic}, are clearly visible -- an energy independent elastic cross section and the $1/v$ law for the inelastic scattering cross section. Sharp features and nonmonotonicity of the scattering as function of $F_2$ and in particular of $\omega$ are present, resembling shape- and Fano-resonances (we return to this point briefly in \seq{Sec:Outlook}). The presence of the pole at $\bar{\omega}^c\approx -1.9995$ suggests that at the critical point, an incoming s-wave with low energy $\omega=\bar{\omega}^c+2$ emits one quantum of energy into the drive and is completely captured into the (long-lived) bound state, only to be radiated as a (mostly) p-wave after absorbing two quanta from the drive.

\subsection{Approximations}\label{Sec:Approximation}

We now discuss the truncation of the potential introduced in \eq{Vin1Vout1} that we repeat here,
\be \begin{array}{cc}
& V_{{\rm in}}^{(1)} \left(\vec{r},t\right)=\left[V_{{\rm int}}\left(r\right)\right]\Theta(d-r),\\\\& V_{{\rm out}}^{(1)} \left(\vec{r},t\right)=\left[-\ddot{\vec{F}}\left(t\right)\cdot \vec{r} +V_{F} \left(t\right)\right]\Theta(r-d),\label{Vin1Vout1b}\end{array}\ee
which corresponds to truncating the axial drive inside the well. The solutions presented in the subsections \ref{Sec:Looselys-wave}-\ref{Sec:p-wave} were all obtained using this form of the potential. The solutions presented in \fig{Fig:OmegaSpace0} do not employ this truncation, but rather solve the full problem with
\be \begin{array}{cc}
& V_{{\rm in}}^{(2)} \left(\vec{r},t\right)=\left[V_{{\rm int}}\left(r\right)-\ddot{\vec{F}}\left(t\right)\cdot \vec{r} +V_{F} \left(t\right)\right]\Theta(d-r),\\\\& V_{{\rm out}}^{(2)} \left(\vec{r},t\right)=\left[-\ddot{\vec{F}}\left(t\right)\cdot \vec{r} +V_{F} \left(t\right)\right]\Theta(r-d),\end{array} \label{Vin2Vout2}
\ee
with the parameters for \fig{Fig:OmegaSpace0} being
\be -A/\pi\approx 0.504, \qquad V_0=1.977.\label{Eq:Fig1params}\ee

\begin{figure}[t!]
\center {\includegraphics[width=3.1in]{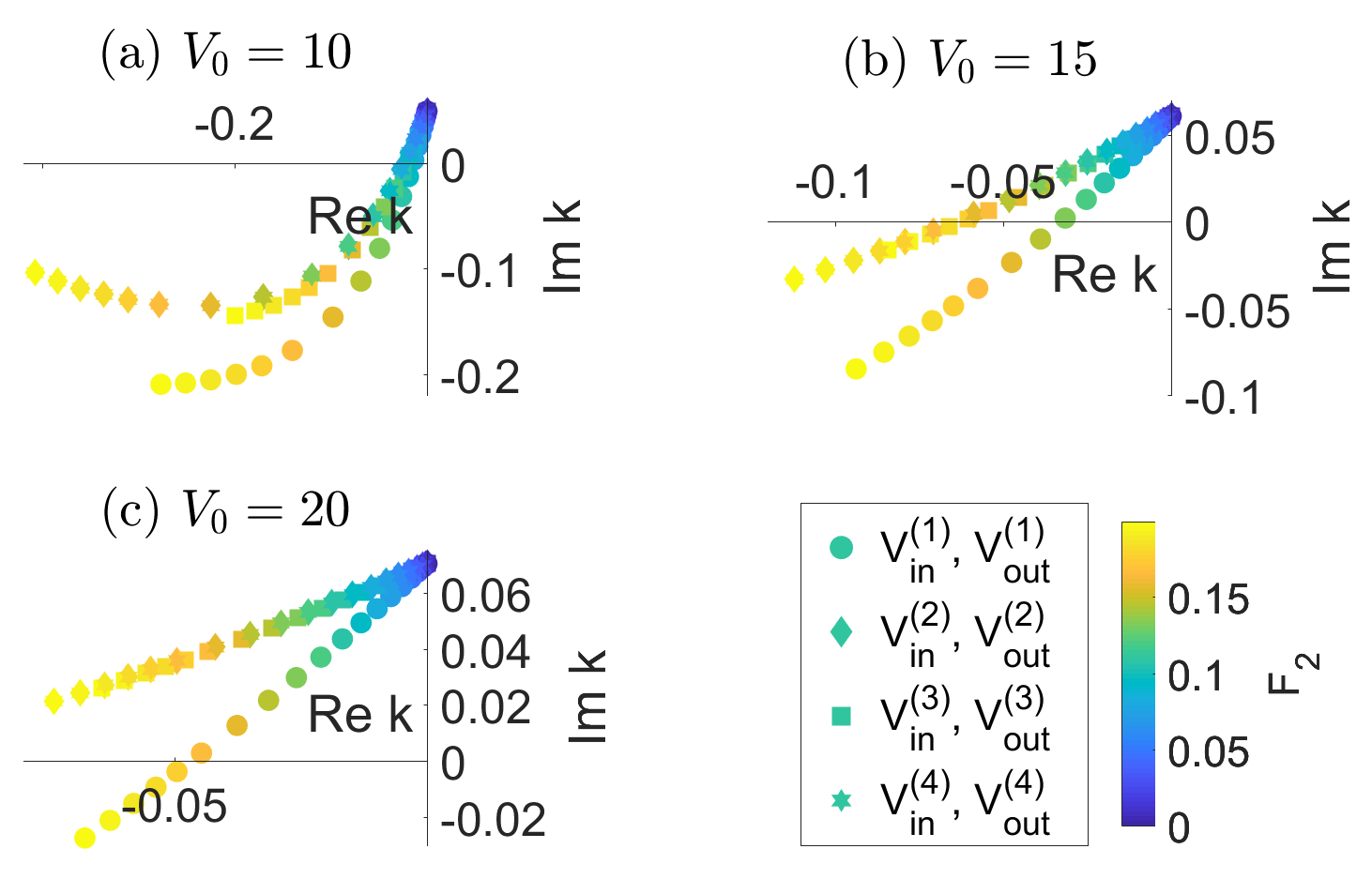}
\caption
{Complex $k$-space comparing poles solved using different potentials defined in \seq{Sec:Approximation}, studying the approximation involved in truncating the drive inside the well, and also the effect of $V_F$ of \eq{Vforce}. See the text for the analysis.
\label{Fig:kSpaceApprox}}}
\end{figure}

For $V_{{\rm in}}^{(2)}$ of \eq{Vin2Vout2}, the wavefunctions in the interior region become time dependent and have to be Fourier expanded as in the exterior region (making the matrices $C$ and $F$ of \eq{eq:matrixform} nondiagonal).
As noted above, $V_F(t)$ is the result of taking \eq{ipsidot} as the physical starting point (an oscillating center of the potential). $V_F$ then cancels the prefactor in \eq{gtint},
\be e^{-i\int^t \frac{1}{2}\dot{F}^2(t')dt'}\label{expF2},\ee
that would otherwise multiply the wavefunction.
 If \eq{iphidot0} is the physical potential (a static potential with an external, periodically modulated linear force), we can define $V_{{\rm in}}^{(3,4)}$ and $V_{{\rm out}}^{(3,4)}$ to be equal to $V_{{\rm in}}^{(1,2)}$ and $V_{{\rm out}}^{(1,2)}$ with $V_F(t)$ set to 0. Then the solutions have to be matched including the term of \eq{expF2} that shifts the quasi-energy and modifies also the Fourier expansion.

\begin{figure}[t!]
\center {\includegraphics[width=3.2in]{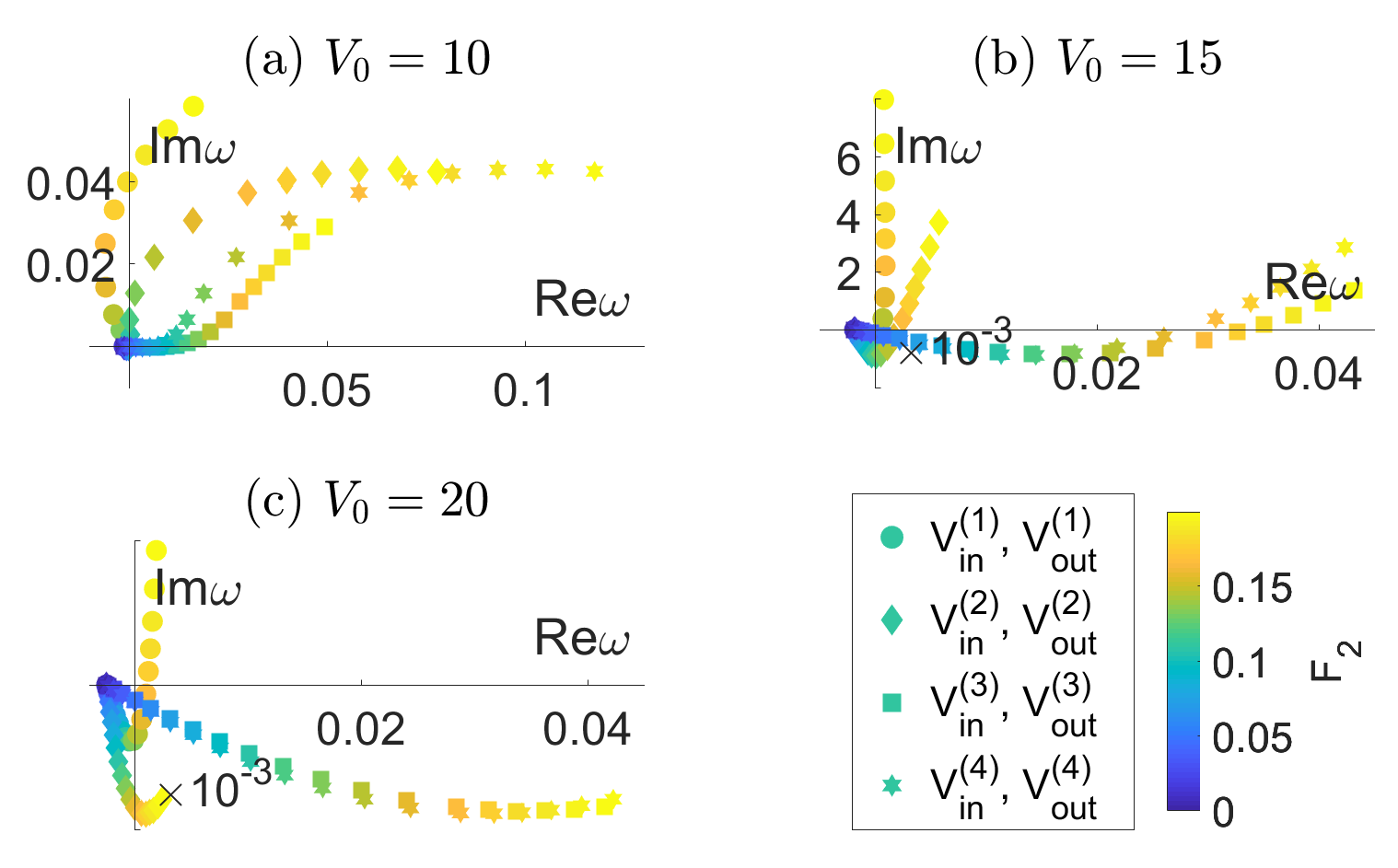}
\caption
{Complex $\omega$-space for the poles of \fig{Fig:kSpaceApprox} solved using different potentials (see text for details). 
\label{Fig:omegaSpaceApprox}}}
\end{figure}

 Figures \ref{Fig:kSpaceApprox}-\ref{Fig:omegaSpaceApprox} show the pole trajectories for a  near-threshold s-wave bound state of a well with $-A/\pi\approx 2.5037$, solved for three increasing different well depth $V_0$, and for each value, comparing the four potentials $V_{{\rm in}}^{(p)}+V_{{\rm out}}^{(p)}$ with $p=1,2,3,4$. 
 
 The most notable result is that the singularities studied in this work exist with all of the studied forms of the potential. They are the result of tunneling and interference, and since they exist with $p=3$, they are not hindered by the possibility to directly couple to the continuum by absorbing energy within the well. The solution with $p=1$ approximates the potential with $p=2$, that can be considered as the `true' potential in a realization that starts from \eq{ipsidot}, although there are noticeable differences in the slopes in momentum space, which lead to some deviations in $\omega$ space as well. This approximation depends on the norm of the wavefunction in the exterior region (which is subject to drive). The poles with $p=3$ and $p=4$ are closer to each other (for deeper well depths) in energy space because of the importance of $V_F$, although in momentum space they do not coincide -- in fact the poles for $p=2$ and $p=4$ coincide exactly in momentum space, because when $V_F$ appears both within and outside the well, its value does not enter the momentum matching.

We note the difference between the expansion presented in the previous sections and the well known Floquet formalism for treating periodic Hamiltonians 
\cite{PhysRev.138.B979,PhysRevA.7.2203,tannor2007introduction}, or time-dependent perturbation theory \cite{RevModPhys.44.602}.
The periodicity of the Hamiltonian allows defining an extended Hilbert space in position and time, that can be spanned by a set of spatially orthogonal wavefunctions and a Fourier basis for time-periodic functions. The current expansion however, employs exact wavefunctions (at least asymptotically), that vary with $\omega$ and are not separable in time and space, in contrast to an expansion using a fixed separable basis that would typically require significantly more basis functions. The connection to the physical solutions is transparent and the explicit use of analytic wavefunctions in each region gives access to details of the spectrum which may be hard to locate otherwise, and in particular Quantum Defect Theory (QDT), discussed in \seq{Sec:Intro}, can be used for the expansion of  wavefunctions in the interior region. Our approach is nonperturbative in both potentials, but neglects the effect of either potential in some region of space and the obtained solutions can be considered, if necessary, as a starting point for an expansion that will correct for the neglected contributions.

Finally we note that the presented expansion and numerical results have been verified by using two different numerical routines in two different programming languages (Matlab and Mathematica), and then directly by plugging the explicit wavefunction into the time-dependent Schr\"{o}dinger equation and verifying the vanishing of both sides of the equation, to the numerical precision possible in the calculation and according to the truncated components.

\begin{figure}[t!]
\center {\includegraphics[width=3.1in]{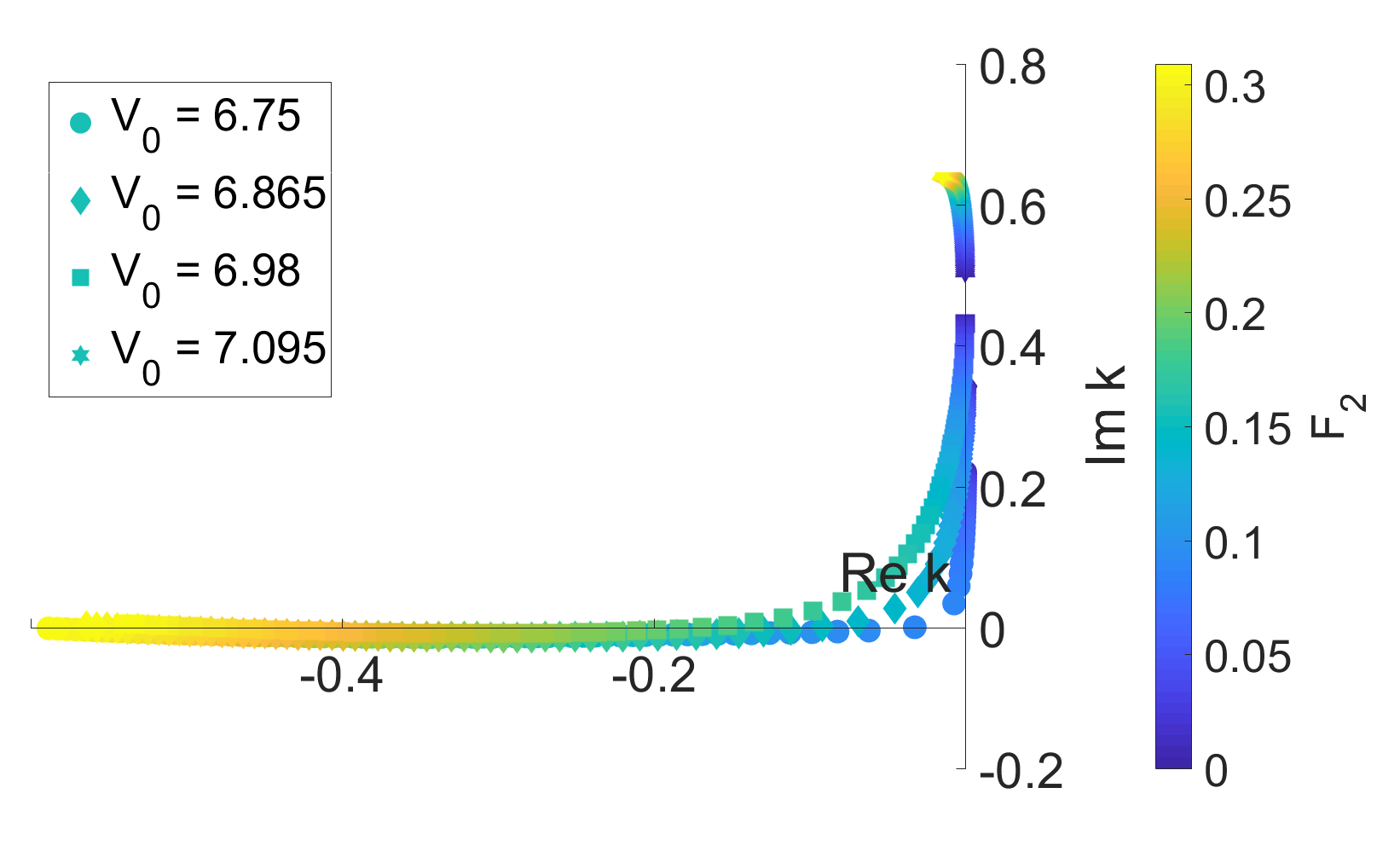}
\caption
{Complex $k$-space showing the value of $k_2$ at the poles of \fig{Fig:OmegaSpace2}. The pole trajectories can be seen to deviate significantly from the almost straight lines seen (for different parameters) in \fig{Fig:kSpace1}. A further study is required in order to explain this and possibly relate it to `resonance interaction' and to the nearby exceptional point (see the text and \fig{Fig:Exceptional}).
\label{Fig:kSpace2}}}
\end{figure}

\section{Outlook}\label{Sec:Outlook}

The momentum space values of $k_2$ for the poles followed in \seq{Sec:p-wave} (\fig{Fig:OmegaSpace2}) are shown in \fig{Fig:kSpace2}. In contrast to \fig{Fig:kSpace1} of \seq{Sec:Looselys-wave}, here the pole trajectories in momentum space deviate significantly from straight lines. This may correspond to the proximity of other poles (`resonance interaction'), and to the existence of the exceptional point nearby \cite{rotter2004influence,heiss2012physics}, discussed below. 

\begin{figure}[t!]
\center {\includegraphics[width=3.1in]{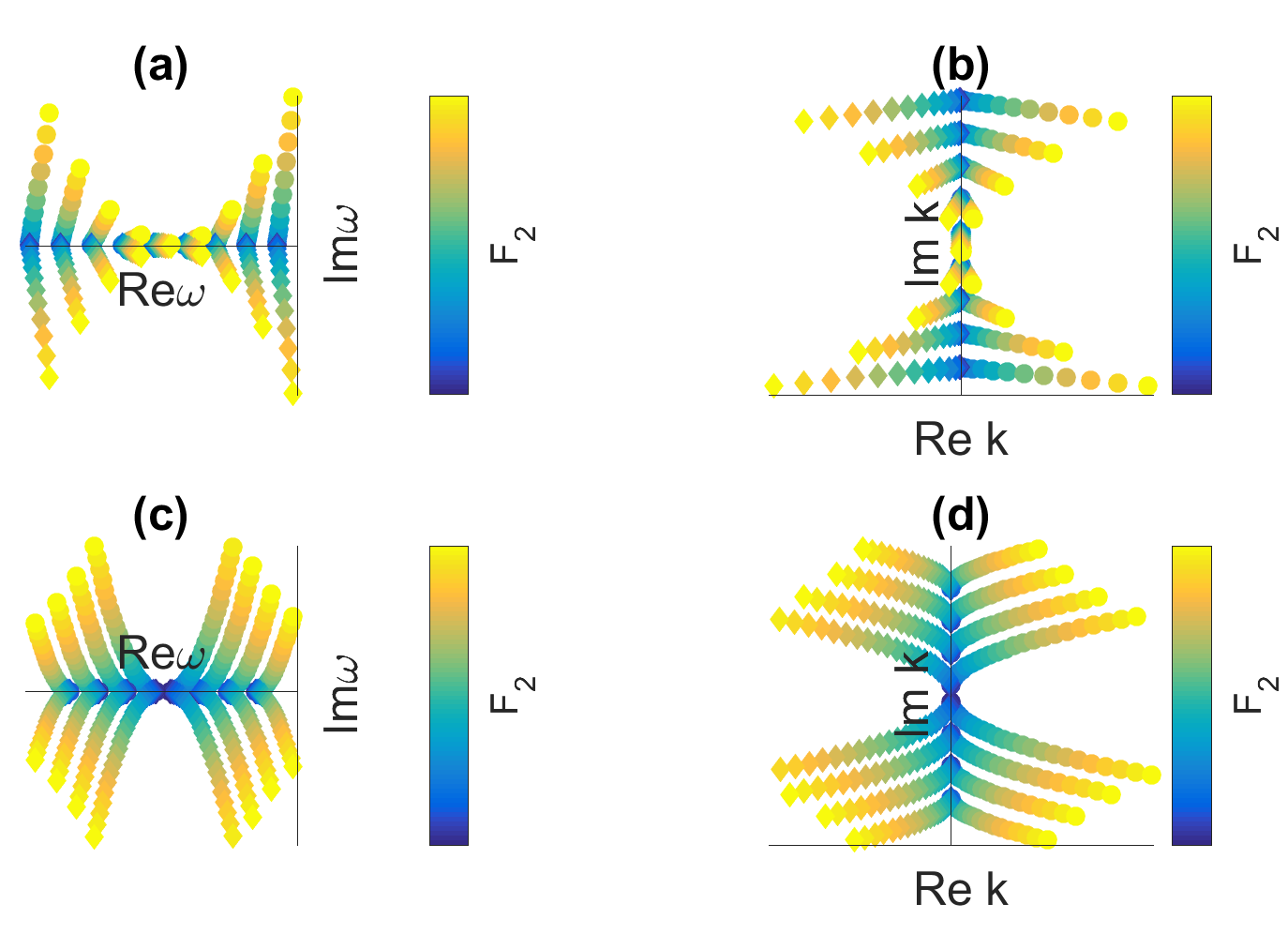}
\caption
{(a) Complex $\omega$-space and (b) complex $k$-plane for a series of poles with a fixed value of $A$, as in \fig{Fig:OmegaSpace1}, and $V_0$ varied between $1.15$ and $1.3$ (the 9 nearly-straight lines starting a adjacent values of $\omega$ or ${\rm Im}k$ respectively). The axis markings and the scale of $F_2\ll 1$ (given by the color code) have been removed for clarity. A smooth rotation of the two pole curves (in both energy and momentum planes) can be seen as function of $V_0$. This is to be contrasted with the seemingly nonanalytic point visible in (c) complex $\omega$-space and (d) complex $k$-plane for a series of p-wave poles with $A$ fixed as in \fig{Fig:OmegaSpace2} and \fig{Fig:kSpace2}, and $V_0\in [7.078,7.091]$. Here it is a Floquet-induced (time-independent) exceptional point, where the p-wave bound state with energy $\omega\approx -2.125$ coincides in energy (mod 2) with the s-wave of $\omega\approx -0.125$, for $V_0\approx 7.09$. A further study is required to confirm a possible line of exceptional points passing through this point.
\label{Fig:Exceptional}}}
\end{figure}

Figure \ref{Fig:Exceptional} compares two scenarios for the pole trajectories in complex momentum and energy planes when varying $V_0$ continuously. In panels (a) and (b), a smooth rotation of the pole trajectories can be seen as their slope changes continuously with $V_0$, going through a point which plausibly shows an interchange of the poles which become the capture and emission poles (that are indistinguishable here at $F_2=0$, but possibly could be distinguished with a Floquet invariant like the Krein signature \cite{Yakubovich}). In panels (c) and (d), $A$ is fixed as in \fig{Fig:kSpace2}, and $V_0$ is varied in small range around the point at which the pole trajectories of \fig{Fig:kSpace2} seem to `branch'. The existence of an exceptional point is clearly seen, where two bound states for $F_2=0$ coincide in energy (mod 2), around which the parametric dependence appears to be nonanalytic. An initial study indicates an entire line of exceptional points that emanates from this point in $(V_0,F_2)$ parameter space. Further study is required in order to check whether an exceptional point can be followed up to the real $\omega$-axis as the Floquet poles studied above, whence it may share further similarities with a spectral singularity of scattering with a real energy. As discussed in \seq{Sec:Intro}, the role of exceptional points in non-Hermitian (open) systems  is attracting increasing attention, and new effects are being actively explored.

The cross sections in \fig{Fig:p-wavecurves} show features resembling shape- and Fano-resonances, which are important in atomic and also nanoscale structures \cite{flach2003fano,miroshnichenko2010fano,ott2013lorentz} (with the distinction that here the scattering is inelastic, and the potential is time-dependent and non-central,  both of those aspects playing an important role). 
As discussed in \seq{Sec:Analytic}, the two linearly indepenednt solutions of a power law potential $-C/r^{\alpha}$ with $\alpha>3$, are analytic in complex energy and momentum planes. Hence, when applying the expansion of the current work to match such wavefunctions in the interior region with the driven particle wavefunctions in the exterior region, no nonanalyticity is introduced in the entire $k$-plane (at any fixed $r\neq d$). Therefore, following the poles of the driven problem in $k$-plane for such a potential should be possible (a-priori) as shown here for a finite range potential. A further study of the possibility to tailor scattering and resonances using Floquet driving, in particular of atomic systems \cite{ZeroFloquet}, is a promising direction to apply the techniques developed in this work. Indeed, the singular points analyzed in the current work can be found in an explicit calculation employing QDT with the polarization interaction ($\alpha=4$) of a cotrapped ion-atom system \cite{unpublished}. More broadly, Floquet driven systems are often analyzed in terms of a time-independent, effective approximation. Our results show a scenario where accounting for the time-dependent nature of the drive is essential, and the implications to many-body Floquet systems \cite{della2013spectral, lindner2011floquet,goldman2014periodically,mitrano2016possible} present an intriguing future direction to explore.

\begin{acknowledgments}
 H.L. thanks Georgy Shlyapnikov, Denis Ullmo, Dmitry Petrov, Andrew Sykes, Pablo Rodriguez, Ido Gilary, Nimrod Moiseyev and Roni Geffen for fruitful discussions, and acknowledges support by the 2013-2014 Chateaubriand fellowship of the French embassy in Israel, support from COST Action MP1001 (Ion Traps for Tomorrow's Applications), through a Short Term Scientific Mission grant, support by a Marie Curie Intra European Fellowship within the 7th European Community Framework Programme, and support by IRS-IQUPS of Universit\'{e} Paris-Saclay.
\end{acknowledgments}

%\clearpage

\begin{widetext}
\appendix

\section{Expansion of linearly-driven cylindrical waves in spherical waves}\label{Sec:Derivations1}

In the following we will use a representation of the spherical Hankel function of the first kind as an integral over cylindrical waves \cite{DanosMaximon,bostrom1991transformation} in the form
\be h_{l}^{\left(1\right)} \left(kr\right)P_{l}^{m} \left(\cos \theta \right)e^{im\varphi } =\int _{C^{(1)}}d\alpha \sin \alpha \frac{1}{2} \left(-i\right)^{l-m} P_{l}^{m} \left(\cos \alpha \right)\chi _{m}^{\left(1\right)} \left(\vec{r};k,\alpha \right)\label{hlexpansion}\ee
where $P_l^m$ are the associated Legendre polynomials and the directed contour of integration $C^{(1)}$ lies in complex $\alpha$-plane. For $k$ with a positive imaginary part we must take $C^{(1)}=\pi /2+i\left(\infty,-\infty\right)$, on which $\cos\alpha \in i\left(\infty,-\infty\right)$, and $\sin\alpha \in \left(0,\infty\right)$. Then $h_{l}^{\left(1\right)} \left(kr\right)$ decays asymptotically as $e^{-({\rm Im}k)r}/r$ (the integral which gives $h_{l}^{\left(1\right)} \left(kr\right)$ is well defined for any $\rho >0$, and decays for $r\to \infty $, which is just what is required for the validity of the solution). For $k$  real and positive the contour of integration is given by $C^{(1)}=i\left(\infty ,0\right)+\left[0,\pi \right]+\left\{\pi +i\left(0,-\infty \right)\right\}$, on which $\cos\alpha \in \left(\infty,-\infty\right)$, and $\rm{Im}\sin\alpha \ge 0$.
For any value of $k$, we have similarly to \eq{hlexpansion}
\be j_{l} \left(kr\right)P_{l}^{m} \left(\cos \theta \right)e^{im\varphi } = \int _{\left[0,\pi \right]}d\alpha \sin \alpha \frac{1}{2} \left(-i\right)^{l-m} P_{l}^{m} \left(\cos \alpha \right)\chi _{m}^{\left(J\right)} \left(\vec{r};k,\alpha \right),\label{jlexpansion}\ee
where $j_l$ is a spherical Bessel function. Writing 
$ h_{l}^{\left(2\right)} =2j_{l} -h_{l}^{\left(1\right)}$,
 and using eq{jlexpansion} and the fact that the $l,m$-dependent coefficients in \eqs{hlexpansion}-\eqref{jlexpansion} are identical, we get an expression identical in form to \eq{hlexpansion}, with the outgoing waves replaced by incoming waves $h_{l}^{\left(2\right)}$, and a contour $C^{(2)}$.
A similar derivation can be repeated in the lower half of complex $k$ plane, making the expansion valid for every complex $k$ (which also follows by analytic continuation).

The proof of \eq{intphiRexpansion} proceeds by using \eq{phiRm12} to write
\be
\begin{array}{l} {e^{i\dot{F}^{\pi } \left(t\right)z} \int_{C^{(a)}} d\alpha \sin \alpha \chi _{m}^{\left(a\right)} \left(\vec{r};k_{2j} ,\alpha \right)b_{{2j} }^{\left(a\right)} \left(\alpha \right)e^{-iF^{\pi } \left(t\right)k_{2j} \cos \alpha }  }
 \\\\\qquad {=e^{i\dot{F}^{\pi } \left(t\right)z} \int_{C^{(a)}} d\alpha \sin \alpha \chi _{m}^{\left(a\right)} \left(\vec{r};k_{2j} ,\alpha \right)\sum _{l_{1} }b_{2j,l_{1} } P_{l_{1} } \left(\cos \alpha \right)\sum _{l_{2} }\left(-i\right)^{l_{2} } \left(2l_{2} +1\right)j_{l_{2} } \left(F^{\pi } \left(t\right)k_{2j} \right)P_{l_{2} } \left(\cos \alpha \right)   } 
\\\\\qquad {=e^{i\dot{F}^{\pi } \left(t\right)z} \sum _{l_{1} ,l_{2} }b_{2j,l_{1}  } \left(-i\right)^{l_{2} } \left(2l_{2} +1\right)j_{l_{2} } \left(F^{\pi } \left(t\right)k_{2j} \right) \sum _{l_{3} }W\left(P_{l_{1} } ,P_{l_{2} } ,P_{l_{3} }^{m} \right)\int_{C^{(a)}} d\alpha \sin \alpha \chi _{m}^{\left(a\right)} \left(\vec{r};k_{2j} ,\alpha \right)P_{l_{3} }^{m} \left(\cos \alpha \right)  } 
\\\\\qquad {=e^{i\dot{F}^{\pi } \left(t\right)z} \sum _{l_{1} ,l_{2} }b_{2j,l_{1}} j_{l_{2} } \left(F^{\pi } \left(t\right)k_{2j} \right)\sum _{l_{3} }c_{l_{1} ,l_{2} ,l_{3} } h_{l_{3} }^{\left(a\right)} \left(k_{2j} r\right)P_{l_{3} }^{m} \left(\cos \theta \right)e^{im\varphi }   } 
\\\\\qquad {=\sum _{l_{4} }i^{l_{4} } \left(2l_{4} +1\right)j_{l_{4} } \left(\dot{F}^{\pi } \left(t\right)r\right)P_{l_{4} } \left(\cos \theta \right)\sum _{l_{1} ,l_{2} }b_{2j,l_{1}} j_{l_{2} } \left(F^{\pi } \left(t\right)k_{2j} \right)\sum _{l_{3} }c_{l_{1} ,l_{2} ,l_{3} } h_{l_{3} }^{\left(a\right)} \left(k_{2j} r\right)P_{l_{3} }^{m} \left(\cos \theta \right)e^{im\varphi }    } 
\\\\\qquad {=\sum _{l_{1} ,l_{2} ,l_{3} ,l_{4} }b_{2j,l_{1}} j_{l_{2} } \left(F^{\pi } \left(t\right)k_{2j} \right)j_{l_{4} } \left(\dot{F}^{\pi } \left(t\right)r\right)h_{l_{3} }^{\left(a\right)} \left(k_{2j} r\right)\sum _{l}c_{l_{1} ,l_{2} ,l_{3} ,l_{4} ,l} Y_{l}^{m} \left(\theta ,\varphi \right)  }
 \end{array}\label{AppEq1}\ee
where the multiplicative factors $e^{-i\frac{1}{2} k_{2j} ^{2} t}$, $N_{2j,l_1}^m$, and $S_{2j,l_1}^{\left(a\right)}$ have been omitted for simplicity, and by using the definition of $R_{2j,l_{1} ,l}^{\left(a\right)} \left(r,t\right)$ given in \eq{R2jl1la}, \eq{AppEq1} results in \eq{intphiRexpansion}. In the derivation of \eq{AppEq1}, the plane-wave expansion in terms of spherical Bessel functions has been used (twice), the coefficients of expansion of a product of two (associated) Legendre polynomials (which can be written using Wigner 3-j symbols) are defined by 
\be W\left(P_{l_{1} }^{m_{1} } ,P_{l_{2} }^{m_{2} } ,P_{l_{3} }^{m_{3} } \right)=\left[2\left(l_3+m_3\right)!/\left(\left(2l_3+1\right)\left(l_3-m_3\right)!\right)\right]^{-1}\int _{-1}^{1}P_{l_{1} }^{m_{1} } \left(w\right)P_{l_{2} }^{m_{2} } \left(w\right)P_{l_{3} }^{m_{3} } \left(w\right)dw,\label{Wdef}\ee
the coefficients $c_{l_{1} ,l_{2} ,l_{3} } $ are obtained using \eq{hlexpansion} and \eq{Wdef} and given by
\be c_{l_{1} ,l_{2} ,l_{3} } =2\left(2l_{2} +1\right)\left(-i\right)^{l_{2}}i^{ l_{3}-m} W\left(P_{l_{1} } ,P_{l_{2} } ,P_{l_{3} }^{m} \right),\ee
and the coefficients $c_{l_{1} ,l_{2} ,l_{3} ,l_{4} ,l} $ are similarly
\be c_{l_{1} ,l_{2} ,l_{3} ,l_{4} ,l} =c_{l_{1} ,l_{2} ,l_{3} } \left(2l_{4} +1\right)i^{l_{4} } W\left(P_{l_{3} }^{m} ,P_{l_{4} } ,P_{l}^{m} \right)/N_{l}^{m},\label{cl1l2l3l4l}\ee
with the definitions
\be Y_{l}^{m} \left(\theta ,\varphi \right)=N_{l}^{m} P_{l}^{m} \left(\cos \theta \right)e^{im\varphi } ,\qquad \qquad N_{l}^{m} =(-1)^m\sqrt{\left(2l+1\right)/4\pi }\sqrt{\left(l-m\right)!/\left(l+m\right)!}.\label{Nlmdefinition}\ee

\section{The expectation value of tensor operators}\label{Sec:Derivations3}

In this appendix we give explicitly the expansion of integrals which are required in order to calculate expectation values of general tensor operators, in the Floquet eigensolutions of \seq{Sec:Matching3D}. For simplicity we treat  here only the most useful case of axially symmetric wavefunctions, with $m=0$ (no $\varphi$ dependence). 
Using the notation of \eq{phi1overruY}, we start by writing the $\pi$-periodic part of the wavefunction in the form
\be \phi ^{\pi } \left(\vec{r}\right) = \sum_{n,l}a_{2n,l}e^{-i2nt}\frac{1}{r}u_{2n,l} \left(r\right) Y_l^0,\label{phipiint} \ee
which corresponds to the expansion in \eq{phi3Dansatz} of wavefunctions in the interior region. For such wavefunctions, we define the (unnormalized) expectation value in the interior region of a purely radial operator $\mathcal{O}\left(r\right)$,
\be \mathfrak{I}_0\left[\mathcal{O}\left(r\right)\right]\equiv\int d^3\vec{r}\left|\phi^{\pi} \left(\vec{r}, t\right)\right|^{2} \mathcal{O}\left(r\right) =  \sum_{\left(n,l\right),\left(n',l'\right)} \delta_{l,l'}e^{2i\left(n-n'\right)t}a_{2n,l}^*a_{2n',l'}\int d r\left[u_{2n,l} \right]^{*}  \mathcal{O}\left(r\right) u_{2n',l'}. \ee
The above expression can be rewritten as
\be \mathfrak{I}_0\left[\mathcal{O}\left(r\right)\right]= \sum_{l}   I_{l,l}\left[\mathcal{O}\left(r\right)\right],\ee
where we have defined for convenience the functional [symmetric under the exchange $\left(n,l\right) \leftrightarrow \left(n',l'\right)$]
\be {I}_{l,l'}\left[\mathcal{O}\left(r\right)\right]=   \sum_{n\leq n'}\left(2-\delta_{n,n'}\right)\,{\rm Re}\left\{e^{2i\left(n-n'\right)t}a_{2n,l}^*a_{2n',l'}\int dr\left[u_{2n,l} \right]^{*}  \mathcal{O}\left(r\right) u_{2n',l'} \right\},\label{Ill}\ee
with the summation taken over pairs of states enumerated by $\left\{\left(n,l\right),\left(n',l'\right)\right\}$ with fixed $l$ and $l'$ obeying $n\leq n'$.

For example, the normalization integral calculated for any time (see \app{Sec:Normalization}) can be written  as
\be \mathfrak{I}_0\left[\hat{1}\right]=\sum_{l} I_{l,l}\left[\hat{1}\right],\label{I1integral}\ee
with $\hat{1}$ the identity operator. Any other expectation value must then be divided by the value of this normalization integral. Similarly, the expectation value of the squared angular momentum operator $\vec{L}^2$ is given by
\be \mathfrak{I}_0\left[\vec{L}^2\right]=\sum_{l} l\left(l+1\right) I_{l,l}\left[\hat{1}\right].\label{L2integral}\ee

For an operator of a general radial part multiplied by the position vector, $\mathcal{O}\left(r\right)\vec{r}$, only the Cartesian $z$-component survives the integral (for axially symmetric wavefunctions), and we can write using $ z/ r=\cos\theta$
\be \mathfrak{I}_1\left[\mathcal{O}\left(r\right)\vec{r}\,\right] = \int d^3\vec{r}\left|\phi^{\pi} \left(\vec{r} ,t\right)\right|^{2} \mathcal{O}\left(r\right)\vec{r} =  \hat{z}\sum_{\left(n,l\right),\left(n',l'\right)}p_{l,l'} e^{2i\left(n-n'\right)t}a_{2n,l}^*a_{2n',l'}\int dr\left[u_{2n,l} \right]^{*}  \mathcal{O}\left(r\right)r\, u_{2n',l'}\,\ee
with the coefficients being
\be p_{l,l'} =2\pi N_{l}^{0}N_{l'}^{0}\int d\theta \sin \theta \cos \theta P_{l} \left(\cos \theta \right)P_{l'} \left(\cos \theta \right).\ee
Using the fact that $p_{l,l'}=p_{l',l}$ and since nonzero terms will have $\left|l-l'\right|=1$, we find
\be \mathfrak{I}_1\left[\mathcal{O}\left(r\right)\vec{r}\,\right] =  \hat{z}\sum_l p_{l,l+1}\left(I_{l,l+1}\left[\mathcal{O}\left(r\right)r\right] +I_{l+1,l}\left[\mathcal{O}\left(r\right)r\right]\right).\ee

For an operator with a general radial part multiplied by a bilinear combination of $\vec{r}$ components, $\mathcal{O}\left(r\right)\vec{r}_{\alpha}\vec{r}_{\beta}$, where $\alpha,\beta\in\left\{x,y,z\right\}$, only the diagonal terms with $\alpha=\beta$ survive the integration (for $m=0$), with the result
\be \mathfrak{I}_2\left[\mathcal{O}\left(r\right)\vec{r}_{\alpha}\vec{r}_{\beta}\right] =\delta_{\alpha,\beta} \sum_{l,l'}q_{\alpha,l,l'} I_{l,l'}\left[\mathcal{O}\left(r\right)r^2\right],\ee
where
\be q_{\alpha,l,l'} =2\pi N_{l}^{0}N_{l'}^{0}\int d\theta \sin \theta \left[ \cos^2 \theta\delta_{\alpha,z}+\frac{1}{2} \sin^2 \theta\left(\delta_{\alpha,x}+\delta_{\alpha,y}\right) \right] P_{l} \left(\cos \theta \right)P_{l'} \left(\cos \theta \right).\ee

In all of the above expressions, ${I}_{l,l'}\left[\mathcal{O}\left(r\right)\right]$ as defined  in \eq{Ill} is valid in the interior region. To get the complete result for expectation values in whole space, the integration over the exterior region must be added, where the wavefunctions are expanded differently in \eq{phi3Dansatz}. In this case, \eq{phipiint} is to be replaced by 
\be \phi ^{\pi } \left(\vec{r} ,t\right) = \sum _{ j,l_{1}} b_{2j,l_{1} }e^{-i2jt}\sum _{l}\frac{1}{r}u_{2j,l_1,l}^{\pi} \left(r,t\right)Y_{l}^0\label{phipiext},\ee
and accordingly, \eq{Ill} becomes in the exterior region
\be {I}_{l,l'}\left[\mathcal{O}\left(r\right)\right]= \sum_{\left(j,l_{1}\right),\left(j' ,l_{1} '\right) }e^{2i\left(j-j'\right)t} b_{2j,l_{1} }^*b_{2j',l_{1}' }
\int dr\left[u_{2j,l_1,l} ^{\pi } \right]^{*}  \mathcal{O}\left(r\right) u_{2j',l_1',l'} ^{\pi },\label{Illext}\ee
with the summation taken over pairs of states enumerated by $\left\{\left(j,l_1,l\right),\left(j',l_1',l'\right)\right\}$ with fixed $l$ and $l'$.
Finally, we note that in the above expressions, the imaginary part of the energy has been omitted -- it gives an exponential envelope of the decay or formation rate of the quasi-bound state. Moreover, all integrals can be performed only on the square-integrable part of the wavefunction, with the nonnormalizable traveling waves omitted from the sums above, in accordance with the interpretation that these belong to the inaccessible part of the Hilbert space.

\section{Normalization of the wavefunction}\label{Sec:Normalization}

The expectation value of any time-independent (or $\pi$-periodic) operator is $\pi$-periodic for the Floquet eigenstates, possibly with an exponential envelope for complex $\omega$.
The normalization integral is not constant in time but rather $\pi$-periodic because the relative weight of the nonnormalizable components oscillates in time (as they are emitted and reflected back during a period of the drive). In order to calculate an expectation value of an operator (determined by the bound components), its integral must be divided by the squared norm, both of which being $\pi$-periodic functions that can be calculated using \app{Sec:Derivations3} (after which averaging is possible).
%We note that also that the integrals in \eq{normalization3D} are a special case of \eq{I1integral}, evaluated at $t=0$. 
The normalization in the interior region can be obtained without explicitly performing the integration, directly from the wavefunctions and their gradients at the matching point. This can be useful especially when the interior wavefunctions are not explicitly known close to the origin, but rather are determined within a QDT formulation \cite{Seaton1983,Gao2008,idziaszek2011multichannel,PhysRevA.84.042703,PhysRevA.87.032706}.
The projection of two eigenfunctions $\phi_1$ and $\phi_2$ of the interior Hamiltonian with energies $\varepsilon_1$ and $\varepsilon_2$ correspondingly, is shown in \app{Sec:Derivations2} to be 
\be 2\,{\rm Re}\int _{0}^{d }\phi _{1} ^{*} \phi _{2} r ^{2} dr = \left(\varepsilon _{1} -\varepsilon _{2} \right)^{-1} {\rm Re}\left. \left\{u_{1} ^{*} u_{2} {{'} } -u_{2} ^{*} u_{1} {{'} } \right\}\right|_{d },\label{phi1phi2int}\ee
where $u_1{'}\equiv \partial _{r} u_{1}$, and  $\varepsilon _{1}$, $\varepsilon _{2}$ are assumed to have equal imaginary parts. The left-hand side of \eq{phi1phi2int} gives the integrals required for the normalization, with the factor of 2 relevant for the off-diagonal projections (when $\phi_1\neq \phi_2$). In the limit of $\phi_1\to \phi_2$ we have for  the diagonal normalization terms
\be \int _{0}^{d }\left|\phi _{1} \right|^{2} r ^{2} dr  = \frac{1}{2}\mathop{\lim }\limits_{\varepsilon _{1} \to \varepsilon _{2} } \left(\varepsilon _{1} -\varepsilon _{2} \right)^{-1} \left. \left[u_{1} ^{*} u_{2} {{'} } -u_{2} ^{*} u_{1} {{'} } \right]\right|_{d}.\ee

%If the wavefunction is square-integrable, the normalization integrals  can be evaluated at $t=0$, and we find using the orthonormality of $Y_{l}^m$, \be 1= \sum _{n,n',l}a_{2n',l}^* a_{2n,l}\int_0^d dr r^2 \left[\phi_{{\rm in},\omega+2n',l}\right]^*  \phi_{{\rm in},\omega+2n,l} +\sum _{\left(j,l_{1}\right),\left(j' ,l_{1} ^{'}\right) }b_{2j',l_{1} ^{{'} } } ^{*} b_{2j,l_{1} } \sum _{l} \int_d^\infty dr r^{2}\left[\phi_{{\rm out},2j',l_{1} ^{{'} } ,l} ^{\pi} \right]^*\phi_{{\rm out},2j,l_{1} ,l }^{\pi}, \label{normalization3D}\ee or in matrix form (using the same indexation used in \eq{eq:matrixform}), \be 1=\vec{a}^{\dag } Q\vec{a}+\vec{b}^{\dag } P\vec{b},\ee so that normalization can be guaranteed by dividing $\vec{a},\vec{b}$ by the square-root of the r.h.s. This normalization is relevant only if the entire wavefunction is square-integrable. 

\section{The projection of two eigenfunctions of the internal Hamiltonian}\label{Sec:Derivations2}

In order to derive \eq{phi1phi2int}, let $\varepsilon _{1} ,\varepsilon _{2} $ be the (possibly complex) energies of two complex eigenfunctions $\phi _{1} ,\phi _{2} $ of the  interior Hamiltonian $H_{\rm in}=-\frac{1}{2}\nabla^2+V_{\rm in}$. For the projection of the two within the interior region, we can write
\be 0=\left\langle \phi _{2} \left|\left(H_{\rm in} -\varepsilon _{1} \right)\left|\phi _{1} \right. \right. \right\rangle -\left\langle \phi _{1} \left|\left(H_{\rm in} -\varepsilon _{2} \right)\left|\phi _{2} \right. \right. \right\rangle.\ee
By canceling the potential energy terms, we get after rearranging the kinetic terms and terminating the integration at an arbitrary point $d$ (which is allowed since the equality above holds identically in space),
\be \begin{array}{l} {\int _{0}^{d }\left(\varepsilon _{1} \phi _{2} ^{*} \phi _{1} -\varepsilon _{2} \phi _{1} ^{*} \phi _{2} \right)r ^{2} dr  =-\frac{1}{2} \int _{0}^{d}\left(u_{2} ^{*} \partial _{r } ^{2} u_{1} -u_{1} ^{*} \partial _{r } ^{2} u_{2} \right)dr  } \\\\ {\qquad =-\frac{1}{2}  \left[\left. \left(u_{2} ^{*} \partial _{r } u_{1} -u_{1} ^{*} \partial _{r} u_{2} \right)\right|_{d } -\int _{0}^{d }\left(\partial_{ r } u_{2} ^{*} \partial _{r} u_{1} -\partial _{r } u_{1} ^{*} \partial _{r } u_{2} \right)dr  \right]} \end{array}\ee
where the factor of $1/2$ is the prefactor in the kinetic energy term $-\frac{1}{2}\nabla^2$, as in \eq{eqradial}. In the second line of the above equation, the integrated term is purely imaginary being the difference of two complex conjugates. Taking the complex conjugate of the entire equation and adding, this term drops and we get
\be \int _{0}^{d }2\,{\rm Re}\left\{\left(\varepsilon _{1} -\varepsilon _{2} \right)\phi _{2} \phi _{1} ^{*} \right\}r ^{2} dr  =-\frac{1}{2}  2\,{\rm Re}\left. \left\{u_{2} \partial _{r } u_{1} ^{*} -u_{1} \partial _{r } u_{2} ^{*} \right\}\right|_{d},\ee
which gives immediately \eq{phi1phi2int}.

\end{widetext}

\bibliographystyle{../hunsrt}

\bibliography{../scattering}

\end{document}